\documentclass[a4paper,11pt,twoside,openright]{thesis}

\usepackage{amsfonts}
\usepackage{amsmath}
\usepackage{graphicx}
\usepackage{amssymb}
\usepackage{esint}
\usepackage{dsfont}
\usepackage[letterpaper]{geometry}
\usepackage{setspace}
\usepackage{stmaryrd}

\DeclareMathAlphabet{\mathpzc}{OT1}{pzc}{m}{it}

\begin{document}

\begin{titlepage}
\begin{center}
September 24, 2010
\\[5cm]
{\LARGE \textbf{Generalized geometry applied to $4d$-supergravity}}
\\[0.5cm]
{\normalsize Under the supervision of Prof. Daniel Waldram}
\\[1.5cm]
{\Large Sara Oriana Tavares}
\\[10cm]
{\normalsize Submitted in partial fulfilment of the requirements \\ for the degree of Master of Science of Imperial College London}
\end{center}
\end{titlepage}

\tableofcontents

\chapter{Introduction} 

Firstly considered on its own as a candidate for a unifying theory of gravity and the electroweak forces, supergravity is currently understood to be the low energy limit of a more fundamental theory \cite{deWitt}. One of the most promising hypothesis to yield this description of quantum gravity is M-theory. Such model can be viewed as the $11d$-theory unifying the five ten-dimensional string-theories, once understood as independent. If we consider this as the picture to emerge at a fundamental level then eleven-dimensional supergravity becomes the most relevant case of study and an important issue is being able to formulate it using a language (hopefully) capable of shedding some light over the underlying theory.

Supergravity is an interesting subject on its own and has been given significant attention since its emergence in the mid-seventies. The simplest case of this extension of general relativity contains only one set ($N=1$) of fermionic generators $\lbrace Q_{\alpha} , \overline{Q}_{\dot{\beta}}\rbrace$ together with the Poincar\'{e} algebra and is developed in four space-time dimensions ($d=4$). The former gives rise to \emph{local} supersymmetry transformations and the associated gauge fields, designated as \emph{gravitinos} $\psi_{\mu}$, together with the metric vielbeins $e^{a}_{\mu}$ are the essential fields present in supergravity. If we consider higher dimensional extensions, gauge fields analogous to the electromagnetic potential, realized as $p$-forms, may emerge, and usual matter fields, such as scalars, can also always be included \cite{Brandt}. From the description above it is apparent we are opting for the Cartan formalism: supergravity is described more naturally in terms of a local frame (provided by the vielbeins) than by making use of a metric. In this picture Lorentz invariance will always be manifest. This is a characteristic of a favourable formalism: one that lets covariance with respect to an important underlying symmetry group naturally appear.  

This is the main idea we will try to reproduce when setting up our language of interest: take advantage of symmetries of M-theory that one might consider to be fundamental and write our supergravity low energy limit in a covariant way with respect to them\footnote{We are using the term symmetry generically to designate an automorphism of a mathematical object, where a morphism is a structure preserving a crucial characteristic of our object.}. The essential step is then to identify the relevant groups that contain such symmetries.

To illustrate the process we shall undertake, let us introduce the well known example of general relativity. Differential geometry is its natural language: on a manifold $M$ of relevance, we will have the vielbeins (and associated metric), connections, covariant derivatives allowing us to set up Lagrangians and derive equations of motion, everything independent of the coordinate system chosen. Underlying all this elegant coordinate-free set up is the symmetry group $Diff(M)$. 

The idea is then to replace $Diff(M)$ by a more interesting group. Without a doubt the features of general relativity should still be present, i.e., the new group should include $Diff(M)$. For $M$-theory two candidates will be of particular interest to us, given their importance in string theory: the continuous groups associated with the $T$-duality and $U$-duality ones, respectively $O(d,d)$ \cite{Grana}, the indefinite orthogonal group of split signature and $E_{d(d)}$ as defined in \cite{Obers}, with $d=\mbox{dim }M$.    

A light description of $T$ and $U$-duality and their significance is due. $T$-duality, for instance, is a symmetry found only in string theories. On the most elementary level it can be viewed as identifying a theory compactified on a circle of radius $R$ with another compactified on one of inverse radius \cite{Aldazabal}. The key issue here is that the momentum modes will be exchanged with winding numbers of the string on the circle, which no longer has an analogue on usual quantum field theories \cite{Aldazabal}. For more evolved compactifications, say on a manifold of dimension $d$, $T$-duality will be realized as the $O(d,d;\mathbb{Z})$ group. 

Say we consider a type II string theory that admits manifolds completely defined by a given metric $g$ as solutions (or, more correctly, as \emph{backgrounds}). Upon $T$-duality the components of this metric will be mixed with those of $B$, an antisymmetric two-form designated as Neveu-Schwarz (NSNS, for short). It certainly seems reasonable trying to find a language in which such exchanges appear rather naturally. As for the $U$-duality group, it consists of an extension of the $T$-duality symmetry, also characteristic of type II string theories \cite{Hull}, that we might consider as the logical ``next step" in a scale of symmetries to include in the structure group.          

Generalized complex geometry, introduced by Hitchin \cite{Hitchin} and developed by Gualtieri \cite{GualtieriPhD} and Cavalcanti, is an example of a powerful formalism to attempt the construction of a language adequate to string theory. With the remarkable property of unifying sympletic and complex manifolds as special cases of a broader structure, a generalized complex manifold, it is proving to be the right tool to use when trying to describe $T$-duality \cite{Cavalcanti} \cite{GualtieriPhD}. The key idea was to look at both geometries as operations in $TM \oplus T^{\ast}M$ instead of just the usual tangent bundle $TM$ \cite{GualtieriPhD}. This new space, doted with a $O(d,d)$ structure group, has a bracket, the \emph{Courant bracket}, with the important feature of admitting \emph{B-transforms} (where $B$ is a closed two-form field) as symmetries as well as diffeomorphisms.  

Despite its richness \cite{GranaFlux}, from a historical point of view the interest in this picture is relatively recent. Manifolds with \emph{fluxes}, i.e., field strengths with a non-vanishing expectation value associated to relevant form-fields such as the NSNS two-form $B$ of closed string theories \cite{Douglas} \cite{Koerber}, were firstly dismissed as appealing string backgrounds for the lack of what were considered essential geometrical properties\footnote{Specifically, they were not K\"{a}hler manifolds.}\cite{GranaFlux}. Only after the emergence of $D$-branes, providing sources for such fluxes \cite{GranaFlux}, did diffeomorphisms and $B$-transforms begin to be regarded on equal footing.       

An extension of Hitchin's generalized geometry, referred to as \emph{exceptional} generalized geometry, was introduced by Hull \cite{Hull} and further studied by Pacheco and Waldram \cite{Waldram}. It consisted in developing a structure where the $O(d,d)$ group of usual generalized geometry is enlarged to the $U$-duality group $E_{d(d)} \supset O(d,d)$ of type II string theories and $M$-theory. One of the ideas presented is of special significance to us, that is, replacing the cotangent bundle $T^{\ast}M$ by $\wedge^2 T^{\ast}M$ in the generalized bundle. In this manner a three-form field $C$, which we encounter to be the object of relevance in $M$-theory along with the metric \cite{Hull}, is encoded instead of the more usual $B$-field.      

In this thesis we will be interested in developing a formalism that drinks from these ideas, but for a toy model of $11d$-supergravity and $M$-theory: ($N=1$, $d=4$) three-form supergravity as introduced by Ovrut and Waldram \cite{Pires}. For such a case, as  it will be seen in the review chapter \ref{ch:membranes}, the relevant field strength will be associated with a three-form and this motivates the analysis of the generalized tangent bundle $TM \oplus \wedge^2 T^{\ast}M$, as already introduced. We will encounter a formalism currently under development by Waldram and his PhD students Coimbra and Strickland-Constable \cite{Coimbra}, doted explicitly with a symmetry group that contains $E_{4(4)} \simeq SL(5,\mathbb{R})$.        

To accomplish the objective of describing this $4d$-supergravity the thesis is roughly divided in two parts: the first and shorter one, comprising chapters two and three, is essentially composed of review work of standard generalized geometry. It tries to familiarize the reader with key properties of the formalism while giving a detailed account of the most relevant concepts for the subsequent study. Following mainly Gualtieri \cite{GualtieriPhD} we present the linear algebra of the generalized tangent bundle and the construction of a generalized metric\footnote{The account for the natural spin structure is provided in Annex \ref{Ap:B}.}. The account for generalized Killing vector fields is based on the work by Gra\~{n}a \emph{et al} \cite{Grana}. The layout chosen will be reproduced when studying $TM \oplus \wedge^2 T^{\ast}M$ in an effort to make comparisons more fruitful. 

The second part begins with a review of three-form supergravity as proposed by Ovrut and Waldram \cite{Pires}, focusing on a simple set up: a purely bosonic action. The description of this model motivates the work on chapters $5$ and $6$: the construction of a $GL(5,\mathbb{R})$ formalism for the generalized tangent bundle, including linear and differential structure, and the introduction of a (pseudo-Riemannian) generalized metric, respectively. These two chapters constitute the core of the thesis original production. Although the majority of the topics covered will be technical, in the sense these chapters are trying to supply us with a rigorous language, sections \ref{sec:Parallel} and \ref{sec:GSusy} bring us back to the original purpose: the description of $4d$-supergravity.       

The ultimate goal, and new to the existing literature, is then writing (maximally symmetric) solutions of our toy model using the specific type of generalized geometry developed and further investigating their properties. The notion of a $p$-parallelizable manifold, generalizing that of manifolds with a nowhere vanishing frame, will play a significant role in bringing supergravity solutions to be seen as flat space. The non trivial example we will focus on is $AdS_4$: a global frame will be constructed and along with it a basis for the (dual) of the generalized tangent bundle. 

In the opinion of the author, throughout this thesis several aspects worth of a deeper investigation were left unstudied. These issues will be summarized in the last chapter as well as a recollection of the main ideas developed. 
   
\chapter{The generalized tangent bundle} \label{ch:1}
 
    Generalized geometry is a tool to supply our theory with a geometrical picture naturally covariant with respect to $B$\emph{-field transforms} (also referred to, in this context, as \emph{gauge transformations}) as well as diffeomorphisms. Suppose a construction involving a manifold $M$ endowed with a metric $g$ and where a closed form field $F$, that can be expressed locally as $F=dA$, plays a significant role (a set up we will develop with three-form supergravity). Now, think of the tangent bundle $TM$. This space is equipped with a (Lie) bracket $\lbrack \cdot , \cdot \rbrack $ such that, for $f \in Diff(M)$ ($X,Y \in TM,q \in M$):
\begin{align}
f_{\ast}(\lbrack X,Y \rbrack {\mid}_q ) = {\lbrack f_{\ast} X,f_{\ast} Y \rbrack {\mid}_{f(q)}}
\end{align}
But note: \emph{only} diffeomorphisms have this property. The fact that we can perform transformations $A \to A+B$, $dB=0$, is not reflected.
    
    Accordingly, our starting point to endow the theory with a new kind of geometry will be to find a space, generalizing $TM$, equipped with a structure, generalizing the Lie bracket, that includes this internal symmetry as well. We will also introduce a new object: a generalized metric that combines the information of both $g$ and $B$.
    
Consider the space $E^p \equiv TM \oplus \wedge^{p} T^{\ast}M$, $p \in \mathbb{N}$, with general element usually denoted as $X + \xi$ or $x$ and referred to as \emph{generalized vector field}. We can define a \emph{generalized Lie derivative}(\cite{Grana}, $p.11$) for this generalized tangent bundle as the Dorfman bracket ($\circ$) (\cite{GualtieriPhD}, $p.25$) over sections on $E$ ($v=V + \zeta, x= X + \xi$):
\begin{align}
\mathbb{L}_v x \equiv (V + \zeta) \circ (X + \xi) = [V,X] + (\mathcal{L}_{V}\xi - i_X d\zeta).
\end{align}
The purpose of the designation \emph{generalized Lie derivative} will become apparent after studying the extension of this bracket to generalized metrics\footnote{See Section \ref{sec:genK}}. Its antisymmetrization gives the \emph{Courant bracket} (\cite{GualtieriPhD}, $p.25$), the one we will treat as the proper generalization of the Lie bracket for vector fields:
\begin{align} \label{def:Courant}
 \lbrack \cdot , \cdot \rbrack \colon & E^p \times E^p \to E^p \nonumber \\  
& (X+\xi,Y+\eta) \mapsto \llbracket X+\xi,Y+\eta \rrbracket = \lbrack X,Y \rbrack 
+ {\mathcal{L}}_X \eta - {\mathcal{L}}_Y \xi - \frac{1}{2} d(i_X \eta - i_Y \xi).
\end{align}

It is not a Lie bracket, on the algebraic sense, for $p \neq 0$, since it fails to satisfy the Jacobi identity. When acting on pure vector fields it reduces to the usual bracket of the tangent bundle as one would expect, and it vanishes for pure forms. One can also prove diffeomorphisms $f \colon M \to M'$ preserve the Courant bracket \cite{GualtieriPhD} ($q \in M$), i.e.:
\begin{align}
f_{\ast}(\llbracket X+ \xi, Y+ \eta\rrbracket){\mid}_q 
= \llbracket f_{\ast} X + (f^{-1})^{\ast} \xi, f_{\ast} Y + (f^{-1})^{\ast}\eta \rrbracket {\mid}_{f(q)}. 
\end{align}

Other transformations preserving this bracket exist but before introducing them, we need to define a new operator on $E^{p}$:
\begin{align}
e^B(X+\xi)\equiv X+(\xi+B(X)),
\end{align}
where we require $\xi+B(X)\in {\wedge}^{p} T^{\ast}M$ and for now $e^B$ is just a notation. One can view $B$ as a $(p+1)$-form via $B(X)=(-1)^{p+1}i_{X}B$, where the factor $(-1)^{p+1}$ is introduced to agree with the usual conventions of the coordinate representation, following \cite{Waldram}. If we now apply this operator to the Courant bracket we will obtain (\cite{GualtieriPhD}, $p.28$): 
\begin{align}
\llbracket e^B(X+\xi),e^B(Y+\eta)\rrbracket & = e^B \llbracket X+\xi,Y+\eta \rrbracket -i_{X}i_{Y}dB \label{eq:CourantB} \\
& = e^B \llbracket X+\xi,Y+\eta \rrbracket \mbox{ iff } dB=0 \nonumber
\end{align}

In fact, if $F$ preserves the Courant bracket, i.e., if $F \llbracket x,y \rrbracket = \llbracket F(x),F(y) \rrbracket $ then $F$ must be a composition of a (push-forward) of a diffeomorphism and a $B$\emph{-field transformation} with $B$ a smooth closed $(p+1)$-form. As mentioned by Sheng (\cite{Sheng}, $p.5$), the proof is identical to the one provided by Gualtieri (\cite{GualtieriPhD}, $p.28$) for the case $p=1$.

However, the introduction of an object combining metric and $B$-field cannot be done so generally: the \emph{generalized metric} $G$ will highly depend on the particular generalized tangent space we are dealing with. The best known example is $E=TM \oplus T^{\ast}M$ and we will use it as the starting point to understand how to construct this new object, hopefully using this knowledge to our advantage when dealing with $E^2=TM \oplus \wedge^{2} T^{\ast}M$, the space, we shall see, indicated for the description of our (specific example of a) $4d$-supergravity. 

\chapter{A starting point: $TM\oplus T^{\ast}M$} \label{sec:Elinear}
The present exposition heavily relies on the work developed by Gualtieri \cite{GualtieriPhD} for generalized complex geometry. The focus is given to the symmetries of $E$ and the construction of a generalized metric. Serving the purpose of completeness, the treatment of its Clifford algebra and spin representations can be found attached\footnote{See Appendix \ref{Ap:B}} but it was not included here given it is not essential for the study we are trying to provide.  
  
\section{Linear structure} 
The space $E$ comes with a natural pairing, to which we will simply refer as the \emph{inner product}, that maps to the smooth functions over $M$ and is usually denoted as follows:
\begin{align}
\langle X+\xi,Y+\eta\rangle = \frac{1}{2}(\eta(X)+\xi(Y)), 
\end{align}
where the notation $\xi(Y)$ makes explicit use of the dual relation between $T^{\ast}M$ and $TM$. In a matrix formulation, regarding the components of $X + \xi $ as a column vector 
$x = {\left( \begin{smallmatrix} X & \xi \end{smallmatrix} \right)}^{tr}$,the inner product can be represented by: 
\begin{align} \label{eq:M}
\langle x,y \rangle = x^{tr}My,
\mbox{  }M = \frac{1}{2} \left( \begin{array}{cc} 0 & \mathds{1} \\ \mathds{1} & 0 \end{array}\right). 
\end{align}
$M$ is a symmetric real matrix so it can be diagonalized by an orthogonal similarity transformation. Moreover, it can be proven it exists $S$ such that $M=S^{tr}DS$ where $D=\mbox{diag}(1,...,1,-1,...-1)$. This means the group preserving the inner product of $E$, $O(E)= \lbrace T \colon T^{tr}MT=M \rbrace $, is isomorphic to $O(d,d)$. Its Lie algebra satisfies:
\begin{align}
\mathfrak{so}(E)& = \lbrace Q \colon Q^{tr}M+MQ = 0 \rbrace, 
\end{align}
or equivalently,
\begin{align}
\mathfrak{so}(E)& =\left\lbrace 
Q \colon Q = \left( \begin{array}{cc} A & \beta \\ B & -A^{tr} \end{array} \right),
{\begin{array}{ccc} 
A \colon & TM \to TM \\
\beta \colon & T^{\ast}M \to TM ,\beta = -{\beta}^{tr} \\
B \colon & TM \to T^{\ast}M, B = -B^{tr} 
\end{array}}
 \right\rbrace. \nonumber
\end{align}
Given the properties of the $B$ field, i.e., since it takes vector fields into one-forms and is antisymmetric, we can regard it as an element of ${\wedge}^2T^{\ast}M$ such that $B(X)=-i_{X}B$. We then see that by choosing $Q_{B} \equiv \left( \begin{smallmatrix} 0 & 0 \\ B & 0 \end{smallmatrix} \right)$, the matrix exponential $e^{Q_{B}}$ will perform the familiar transformation:
\begin{align} 
e^{Q_B}(X+\xi)
= \left( \begin{array}{cc} \mathds{1} & 0 \\ B & \mathds{1} \end{array} \right) 
\left( \begin{array}{c} X \\ \xi \end{array} \right)
= X + (\xi - i_{X}B).
\end{align}
We will abuse the notation slightly and refer to this operator simply as $e^{B}$, matching the notation introduced in the beginning of the chapter. We refer to it as a \emph{shear transformation in the} $T^{\ast}M$ \emph{direction} or, as before, simply as a $B$\emph{-field transformation}. In a much similar fashion we can view $\beta$ as a bi-vector in $\wedge^{2}TM$ via $\beta(\xi)=-i_{\xi}\beta$ (remembering $i_{\xi} \colon \wedge^{p+1} TM \to \wedge^p TM$) and the operator $e^{Q_{\beta}}$ brings us the \emph{ shear transformation in the} $TM$ \emph{direction} $e^{Q_{\beta}}(X+\xi)=(X-i_{\xi}\beta)+\xi$. 

Finally, by considering $Q_{A}= \left( \begin{smallmatrix} A & 0 \\ 0 & -A^{\ast} \end{smallmatrix} \right)$ and noting that any transformation $V$ on the identity component of the linear transformations of $TM$, $GL^{+}(TM)$, can be written as $V=e^{A}$, for some $A \in T^{\ast}M \otimes TM = \mbox{End }(TM)$, we see that by means of $e^{Q_{A}}$ we can regard $GL^{+}(TM)$ has a subgroup of $SO(E)$. As mentioned by Gualtieri \cite{GualtieriPhD}, this diagonal embedding can be extended to include the whole of $GL(TM)$ in $O(E)$ proving the usual linear transformations over $TM$ are not lost when considering the generalization of the bundle. 

We note the decomposition $\mathfrak{so}(E)=\wedge^2 TM \oplus \mbox{End } TM \oplus \wedge^2 T^{\ast}M$ is in agreement with the fact that for a vector space $V$ endowed with a non-degenerate bilinear form it holds $\mathfrak{so}(V)=\wedge^2 V$. 

\section{Generalized metric} \label{sec:GenMetricT1}

To introduce the generalized metric we need to introduce a new partition on $E$. Let $C_{+} \subset E$ be a maximal subspace - a subspace not strictly contained in any other - where the pairing $\langle \cdot ,\cdot \rangle$ is positive definite (it can be proven $\mbox{dim }C_{+}=\mbox{dim }TM=d$), and let $C_{-}$ be its orthogonal complement: $C_{-} \equiv C_{+}^{\perp}$. The inner product will be negative definite in $C_{-}$ so the splitting $E=C_{+} \oplus C_{-}$ determines a metric 
$G \equiv {\langle \cdot ,\cdot \rangle}{\mid}_{C_{+}}-{\langle \cdot ,\cdot \rangle}{\mid}_{C_{-}}$ that can also be seen as an operator:
\begin{align}
G \colon & E = C_{+} \oplus C_{-} \to E^{\ast} \simeq E \nonumber \\
& x = x_{+} + x_{-} \mapsto Gx = {\langle x ,\cdot \rangle}{\mid}_{C_{+}}-{\langle x ,\cdot \rangle}{\mid}_{C_{-}} \simeq x_{+} - x_{-},
\end{align}
making explicit use of the relation $E^{\ast}={(TM \oplus T^{\ast}M)}^{\ast}=(T^{\ast}M \oplus TM)=E$. This is an operator also satisfying
\begin{align}
\langle Gx,Gy\rangle & = \langle x_{+} + x_{-},y_{+} + y_{-} \rangle \nonumber \\
& = \langle x_{+},y_{+}\rangle - \langle x_{+},y_{-}\rangle - \langle x_{-},y_{+}\rangle + \langle x_{-},y_{-}\rangle \nonumber \\
& = \langle x,y \rangle ,
\end{align}
since if $x_{\pm} \in C_{\pm}$ then $\langle x_{+} , x_{-} \rangle = 0$.

The operator $G$ is a symmetric automorphism of $E$ that squares to the identity. We note also $C_{+}$ and $C_{-}$ are its eigenspaces with eigenvalues $+1$ and $-1$, respectively, so, as a metric, $G$ has a split signature $(d,d)$.

$G$ will be our generalized metric but how can we make it apparent? The answer is to find the partition $C_{+} \oplus C_{-}$ explicitly, by making use of a map $\gamma \colon TM \to T^{\ast}M$ satisfying the requirement
\begin{align} 
& \langle X + \gamma(X), X + \gamma(X)\rangle > 0,\mbox{ } \forall_{X \in TM}.
\end{align} 
The splitting is found trivially just by setting
\begin{align}
C_{+}=\Gamma_{\gamma} \equiv \lbrace X + \gamma(X) \colon X \in TM \rbrace.
\end{align}

Taking into account $\gamma$ takes a vector field into a one-form we can regard it as an element of $T^{\ast}M \otimes T^{\ast}M$. The isomorphism $T^{\ast}M \otimes T^{\ast}M \simeq S^{2}T^{\ast}M \oplus {\wedge}^{2}T^{\ast}M$ allows us to set $\gamma \equiv g + B$ and, once again, $B(X)$ is defined via $B(X)=-i_{X}B$. Note how the nil-potency of $i_{X}$ is related to the non-contribution of $B(X)$ to the value of $\langle X + \gamma(X), X + \gamma(X)\rangle$ making the defining condition equivalent to require the symmetric part of $\gamma$, $g$, to define a Riemannian metric on $TM$. One also finds easily the identity $\Gamma_{g+B}=e^{B}\Gamma_{g}$. Finally, if $C_{+}=e^{B}\Gamma_{g}$ then $C_{-}=e^{B}\Gamma_{-g}$.

All this information we have gathered allows us to find the generalized metric in a simple format. The simplest set up to start with is given by the case $B=0$, where $C_{+} = \lbrace X+g(X) \rbrace $ and $C_{-}=\lbrace X-g(X) \rbrace $. By defining $x_{\pm}=X \pm g(X)$ and noting $2X = x_{+} + x_{-}$, $2g(X) = x_{+} - x_{-}$ \cite{Baraglia} the action of our generalized metric, call it $G_{0}$, is easy enough to compute:
\begin{align}
G_{0}(2X)& = x_{+} - x_{-} = 2g(X) \\
G_{0}(2g(X))& = G_{0}(x_{+} - x_{-}) = 2X.
\end{align}
If we then use our vector representation for $X+g(X)$ we find:
\begin{align} 
G_{0}= \left( \begin{array}{cc} 0 & g^{-1} \\ g & 0 \end{array} \right) .
\end{align}  
For $B \neq 0$, calling our generalized metric $G_{B}$, we have ($C_{+}=e^{B}\Gamma_{g}$)
\begin{align}
& G_{B}C_{+} = C_{+}
 \Leftrightarrow(e^{-B}G_{B}e^{B})\Gamma_{g}=\Gamma_{g} \nonumber \\
& \Rightarrow e^{-B}G_{B}e^{B} = G_{0}
 \Leftrightarrow G_{B}= e^{B}G_{0}e^{-B}.
\end{align}
In the matrix formulation we obtain:
\begin{align}
G = \left( \begin{array}{cc} \mathds{1} & 0 \\ B & \mathds{1} \end{array} \right)
\left( \begin{array}{cc} 0 & g^{-1} \\ g & 0 \end{array} \right) 
\left( \begin{array}{cc} \mathds{1} & 0 \\ -B & \mathds{1} \end{array} \right)
=\left( \begin{array}{cc} -g^{-1}B & g^{-1} \\ g -Bg^{-1}B & Bg^{-1} \end{array} \right).
\end{align}
In this case there was no additional restrictions on $B$, but to have $e^{B}$ preserving the Courant bracket we know $B$ must be closed. One can also prove (\cite{GualtieriPhD}, \cite{GualtieriLN}) $g - Bg^{-1}B$ is still a Riemannian metric in $TM$.

It is sometimes more natural to regard the object $H \equiv 2MG$ as the generalized metric \cite{Hull}, where $M$ is the matrix representing the inner product introduced earlier\footnote{Recall equation \eqref{eq:M}}. This transformation brings $G_0$ to a diagonal form (\cite{Grana}, $p.6$). It is perhaps more natural to think of the matrix representation $G$ as $G^m_n$ and of $H$ as $H_{mn}$, $m,n=1,\cdots,2d$ (\cite{Hull}, $p.6$). It will be a Riemannian generalized metric: the signature is $(2d,0)$: 
\begin{align}
H_{0} &= \left( \begin{array}{cc} g & 0 \\ 0 & g^{-1} \end{array} \right) 
\end{align}
\begin{align}
H_B &=\left( \begin{array}{cc} g -Bg^{-1}B  & Bg^{-1} \\ -g^{-1}B  & g^{-1} \end{array} \right). \label{eq:H_B}
\end{align}     

\section{Generalized Killing vector fields} \label{sec:genK}

First we shall study the generalized Lie derivative a little further. Let $x_{\pm} \in C_{\pm}$. Let us think of the conditions $v$ must satisfy for $\mathbb{L}_v(x_{\pm})$ to still have positive or negative definite inner product, respectively. For starters, let us take the case $B=0$. The condition mentioned is then equivalent to:
\begin{align}
& \mathbb{L}_v \left( X \pm g(X) \right) =[V,X] \pm g([V,X]) \\
& \Leftrightarrow \nabla V = dV \mbox{  } \wedge \mbox{  } d\zeta=0 \nonumber , 
\end{align}
with the covariant derivative constructed with the usual metric connection. For $B \neq 0$ we find a similar relation. The only difference is that now $\zeta$ must satisfy $\partial_{[\mu} \zeta_{\nu]}=-\frac{1}{2}V^{\rho}(\partial_{\rho}B_{\mu \nu})$. Remembering the definition of a \emph{Killing vector field} as $V$ satisfying $\mathcal{L}_V g =0$ we see that the condition above defines a possible generalization of this notion, when a $B$-field is present, given the condition $\nabla V=dV$ is equivalent to the one stated.

However, when dealing with a background (\cite{Grana}, $p.10$) where we can perform gauge transformations $B \to B + d\omega$ satisfying $\mathcal{L}_B + d\zeta=0, \zeta \equiv i_Vd\omega$ we would be more interested in defining a generalized killing vector by encoding this information too. 

Recalling the usual Lie derivative when acting on $(2,0)$-tensor fields $T$ satisfies
\begin{align}
(\mathcal{L}_V T)(X,Y)=\mathcal{L}_V(T(X,Y)) - T(\mathcal{L}_V X,Y) - T(X, \mathcal{L}_V Y),
\end{align}
we will define our generalized Lie derivative to act on generalized metrics in analogy (\cite{Grana}, $p.11$):
\begin{align}
(\mathbb{L}_v G)(x,y) \equiv \mathbb{L}_v(G(x,y)) - G(\mathbb{L}_v x,y) - G(x, \mathbb{L}_v y).
\end{align}
and we note $\mathbb{L}_v(G(x,y))= V[G(x,y)]$. Taking $H_B$, previously introduced \eqref{eq:H_B}, the calculation holds:  
\begin{align}
(\mathbb{L}_{v}H_B)(x,y) = & X[\mathcal{L}_{V}(g-Bg^{-1}B)-2 d\zeta g^{-1}B-Bg^{-1}2 d\zeta ]Y  + X[\mathcal{L}_{V}(Bg^{-1})+2d\zeta g^{-1}]\eta \\
& + \xi [-\mathcal{L}_{V}(g^{-1}B)-g^{-1}2d\zeta ]Y  +\xi[\mathcal{L}_{V}g^{-1}]\eta \nonumber
\end{align}
\begin{align}
\Rightarrow
\mathbb{L}_{v}H_B = &  
\left( 
\begin{array}{cc}
\begin{array}{c}
\mathcal{L}_{V}g-(\mathcal{L}_{V}B+2d\zeta)g^{-1}B-B(\mathcal{L}_{V}g^{-1})B \\
-Bg^{-1}(\mathcal{L}_{V}B+2d\zeta)
\end{array} & (\mathcal{L}_{V}B+2d\zeta)g^{-1}+B(\mathcal{L}_{V}g^{-1})\\
-g^{-1}(\mathcal{L}_{V}B+2d\zeta)-(\mathcal{L}_{V}g^{-1})B & \mathcal{L}_{V}g^{-1}
\end{array}
\right).
\end{align}  
If we take $2\zeta \to \zeta$ we verify the conditions we wanted to satisfy have now combined to one. A \emph{generalized Killing vector field} $v$ is then defined to be a vector field satisfying the relation:
\begin{align} \label{eq:Killing}
\mathbb{L}_v H_B = 0.
\end{align}

\chapter{Review work from \emph{Membranes and Three-form supergravity}} \label{ch:membranes}

Instead of treating the usual minimal supergravity action we use a dualised version of it, by replacing one of the auxiliary fields with the help of a four-form field-strength (Ovrut and Waldram, \cite{Waldram}). This action is most relevant as a toy model for an eleven dimensional supergravity but here we shall use it as the starting point theory to describe, using generalized geometry.

The original $N=1$, $d=4$ supergravity (bosonic) lagrangian $\mathcal{L}$ is given by (\cite{Waldram}, $p.19$):
\begin{align}
k^{2}e^{-1} \mathcal{L}= \frac{1}{2}R + \frac{1}{3}b^{2}-\frac{1}{3}M^{2}_1 - \frac{1}{3}M^{2}_2 ,
\end{align}
where $M_1$ and $M_2$ are scalar auxiliary fields, $b_{a}$ a vector one and $R$ is the Ricci scalar ($e$ denotes the determinant of the vielbein $e_{\mu}^{a}$ satisfying $g_{\mu \nu}=\eta_{ab}e_{\mu}^{a}e_{\nu}^{b}$ and $k$ denotes a constant factor). The equations of motion for the auxiliary scalar fields $M_{i}$, $i=1,2$ and vector $b_{a}$ are given by:
\begin{align}
\frac{\partial \mathcal{L}}{\partial M_{i}} - \partial_{\mu} \frac{\partial \mathcal{L}}{\partial (\partial_{\mu} M_{i})} & = 0 \Leftrightarrow -\frac{2}{3} M_{i} - 0 = 0 \Leftrightarrow M_{i} = 0 \\
\frac{\partial \mathcal{L}}{\partial b_{a}} - \partial_{\mu} \frac{\partial \mathcal{L}}{\partial (\partial_{\mu} b_{a})} & = 0 \Leftrightarrow \frac{2}{3} b^{a} - 0 = 0 \Leftrightarrow b_{a} = 0.
\end{align}
So, all the auxiliary fields should vanish. We dualise $M_1$ by setting $M_1=\frac{1}{4}\varepsilon^{\mu_1 \cdots \mu_4}F_{\mu_1 \cdots \mu_4}$, noting that we can always choose 
\begin{align} \label{eq:Cform}
F_{\mu_1 \cdots \mu_4}&=4\partial_{\mu_1}C_{\mu_2 \mu_3 \mu_4},
\end{align}
at least locally, since a four-form in four dimensions is always closed (the constant $4$ is just a normalization factor introduced to agree with the definitions gave in \cite{Waldram}). Note also that $F=dC$ is determined up to a \emph{gauge transformation} $C \to C+B$, where $B$ is a closed form.

While still having $M_{2}=b_{a}=0$, the $M_1$ term leads to a different e.o.m. (the two formalisms are not equivalent). We shall rewrite the Lagrangian in terms of $F$:
\begin{align}
k^{2}e^{-1} \mathcal{L} & = \frac{1}{2}R + \frac{1}{3}b^{2}-\frac{1}{3}M^{2}_1 - \frac{1}{3}{\left( \frac{1}{4}\epsilon^{\mu_1 \cdots \mu_4}F_{\mu_1 \cdots \mu_4} \right)}^2 \nonumber \\
& = \frac{1}{2}R + \frac{1}{3}b^{2}-\frac{1}{3}M^{2}_1 + \frac{1}{2}F^{2}.
\end{align}
Now we can determine the equation of motion for $C$:
\begin{align}
\frac{\partial \mathcal{L}}{\partial C_{\mu_2 \mu_3 \mu_4}} - \partial_{\mu_1} \frac{\partial \mathcal{L}}{\partial (\partial_{\mu_1} C_{\mu_2 \mu_3 \mu_4})} & = 0 \nonumber \\
\Leftrightarrow \partial_{\mu_1}F^{\mu_1 \cdots \mu_4} & = 0 \nonumber \\
\Leftrightarrow \nabla_{\mu_1}F^{\mu_1 \cdots \mu_4} & = 0.
\end{align}
We have found that the field strength $F^{\mu_1 \cdots \mu_4}$ is covariantly constant. Since it is totally antisymmetric it should be proportional to the Levi-Civitta tensor $\varepsilon^{\mu_1 \cdots \mu_4}$. Calling the proportionality constant $\lambda$ we have $F^{\mu_1 \cdots \mu_4}=\lambda \varepsilon^{\mu_1 \cdots \mu_4}$. This leads to:
\begin{align}
M_{2} = \frac{1}{4} \varepsilon^{\mu_1 \cdots \mu_4} (\lambda \varepsilon_{\mu_1 \cdots \mu_4}) = -6 \lambda,
\end{align}
so the auxiliary field as acquired a (in general) non-vanishing expectation value connected to the introduction of the \emph{flux} $F$.
Now, working on shell, where $b_{a}=M_2=0$, we can write our action as:
\begin{align}
S = \frac{1}{2k^{2}} \int d^{4}x \sqrt{|g|}(R+\alpha F^{2}),
\end{align}
where the constant $\alpha > 0$ has been introduced so that the metric e.o.m. can be presented in the simplest way possible. Varying the action with respect to the metric, we have:
\begin{align}
2k^2 \delta S = \delta (\sqrt{|g|})(R+\alpha F^{2})+ \sqrt{|g|} \delta g^{\mu \nu} R_{\mu \nu} + \sqrt{|g|} g^{\mu \nu} \delta R_{\mu \nu} + 2 \alpha \sqrt{|g|}\delta F^{\mu_1 \cdots \mu_4} F_{\mu_1 \cdots \mu_4}.
\end{align}
Let us analyse each variation individually. From the identities $\frac{\delta g}{g} = g^{\mu \nu }\delta g_{\mu \nu}$ and $g^{\mu \nu} \delta g_{\mu \nu} = -g_{\mu \nu} \delta g^{\mu \nu}$ one arrives at $\delta (\sqrt{|g|}) = - \frac{1}{2}\sqrt{|g|}g_{\mu \nu} \delta g^{\mu \nu}$. The term including $\delta R_{\mu \nu}$ is a total divergence, $\sqrt{|g|} g^{\mu \nu}\delta R_{\mu \nu} = \sqrt{|g|} ( \nabla_{\rho}(g^{\mu \nu}\delta \Gamma_{\mu \nu}^{\rho})-\nabla_{\nu}(g^{\mu \nu}\delta \Gamma_{\mu \rho}^{\rho}))$, so by Stoke's theorem it can be integrated into a surface term we assume disappears. Expressing the variation of $F^{\mu_1 \cdots \mu_4}$ as $\delta F^{\mu_1 \cdots \mu_4} = \delta g^{\mu \nu} \frac{\delta F^{\mu_1 \cdots \mu_4}}{\delta g^{\mu \nu}}$, we find $\delta F^{\mu_1 \cdots \mu_4} F_{\mu_1 \cdots \mu_4} = 4F_{\mu}^{\mu_2 \mu_3 \mu_4}F_{\nu \mu_2 \mu_3 \mu_4} \delta g^{\mu \nu}$. Requiring $\delta S = 0$ is then equivalent to set:
\begin{align}
& - \frac{1}{2}g_{\mu \nu}(R + \alpha F^{2})+ R_{\mu \nu} + 8 \alpha F_{\mu}^{\mu_1 \mu_2 \mu_3}F_{\nu \mu_2 \mu_3 \mu_4} = 0 \nonumber \\
\Leftrightarrow & R_{\mu \nu}-\frac{1}{2}g_{\mu \nu}R = 36 \alpha \lambda^{2}g_{\mu \nu},
\end{align}
indicating we should choose $\alpha = \frac{1}{36}$. This equation can yet be rewritten as:
\begin{align} \label{eq:metric}	
R_{\mu \nu}=-\lambda^{2}g_{\mu \nu}.
\end{align}

Although this model is very simple, it is a good starting point to rely on for our construction of generalized geometry. It presents a $3$-form field as the relevant form associated to the flux $F$, which points us to study the generalized tangent bundle $TM \oplus \wedge^2 T^{\ast}M$, as seen in chapter \ref{ch:1}. This will be our project for the next one.  

We will finally be interested in describing the two maximally symmetric solutions (\cite{Petersen}, $p.2$) satisfying the relation \eqref{eq:metric}: Minkowski and Anti de Sitter space for $\lambda = 0$ and $\lambda \neq 0$, respectively. This will be accomplished with the introduction of the generalized metric, that will occupy us throughout chapter \ref{ch:metric}.  

\chapter{Studying $TM\oplus \wedge^{2}T^{\ast}M$}

Over the preceding chapters we have acquired some familiarity with the linear algebra of $E=TM \oplus T^{\ast}M$ and we have found the motivation to study $E^2=TM \oplus \wedge^{2}T^{\ast}M$. Although we will rely on much of the study done for $E$, the two bundles have a fundamental difference: there is no natural bilinear pairing on $E^2$ meaning we no longer have an inner product preserving group of transformations. 

This gives rise to the important question of what group $G$ should be considered to encode the linear structure of $E^2$. Relying on our ultimate goal of constructing a language appropriate to deal with a more fundamental theory than supergravity itself, the idea of including the $U$-duality symmetry as a subgroup comes to form. Given it is possible to embed $GL(4,\mathbb{R})$, the natural group of transformations of $TM$ for $\mbox{dim }M=4$, directly in $E_{4(4)} \simeq SL(5,\mathbb{R})$ setting $G=SL(5,\mathbb{R})$ seems to be reasonable. We, however, settled for choosing $GL(E^2)\equiv GL(5,\mathbb{R})$ and the detailed account for this choice is summarized in the first section.

Having $GL(5,\mathbb{R})$ as the symmetry group makes it convenient to use a different representation for the element $x=X+\xi \in TM \oplus \wedge^{2}T^{\ast}M$. We shall rewrite the Courant bracket in terms of it, giving special attention to the expected symmetries. Finally, a differential structure for the bundle is also introduced.    

\section{Linear algebra of the general element} \label{sec:TMlinear}

We will find the linear structure for the space $E^2$ motivated by the transformations we have found for $SO(E)$ and by considering a $GL(5,\mathbb{R})$ matrix representation\footnote{For a detailed account of the isomorphism giving rise to this representation see Appendix B} of such an element $x$ with entries $x^{mn}$, $m,n=1,...,5$ satisfying ($\mu,\nu_i,\mu_i = 1,\cdots,4$): 
\begin{align}
x^{\nu_1 \nu_2} & = \frac{1}{2} \epsilon^{\nu_1 \nu_2 \mu_1 \mu_2} \xi_{\mu_1 \mu_2} \Rightarrow x^{\nu_1 \nu_2} \in (\det TM)\otimes \wedge^{2}T^{\ast}M\\
x^{5 \mu}& = X^{\mu} \\
x^{mn}&=-x^{nm} \nonumber,
\end{align}
where $\frac{1}{2}\xi_{\mu_1 \mu_2}$ and $X^{\mu}$ denote the components of a two-form and a vector field, respectively, as customary. 

We would like to encounter the general form of $Q \in GL(E^2)$ such that $x'=QxQ^{tr}$ reduces adequately to the known transformations for $TM$ and $\wedge^{2}T^{\ast}M$, which is to say, to find the transformations that include as a subgroup $GL(TM)$. This brings us to a linear transformation that must take a block diagonal form:
\[Q_V= \left( \begin{array}{cc} V(\det V)^{q} & 0 \\ 0 & (\det V)^{p} \end{array} \right),
V \in GL(TM),p,q \in \mathbb{R}. \]
Starting with the transformation for $X^{\mu}$, we get $x^{5 \mu'}={(Q_V)}_{\nu}^{5}{(Q_V)}_{\mu}^{\mu'}x^{\nu \mu}=(\det V)^{q+p}V_{\mu}^{\mu'}x^{5\mu}$. With the usual group of transformations of the vectors being $GL(TM)$ we see that to preserve this symmetry we should set $q+p=0$. Now, for the case of two-forms we have $x^{\mu_1'\mu_2'}={(Q_V)}_{\mu_1}^{\mu_1'}{(Q_V)}_{\mu_2}^{\mu_2'}x^{\mu_1 \mu_2}=V_{\mu_1}^{\mu_1'} V_{\mu_2}^{\mu_2'}(\det V)^{2q}x^{\mu_1 \mu_2} \Rightarrow q =\frac{1}{2}$. 

Note the requirement for $x^{\mu_1 \mu_2}$ and $x^{5\mu}$ to transform has form (density) and vector, respectively, is a very restrictive one, and incompatible with the condition $\det Q_V=1$: imposing the two conditions simultaneously would lead us to $V \in SL(TM)$. This is the reason to set $GL(E^2)=GL(5,\mathbb{R})$ instead of $GL(E^2)=SL(5,\mathbb{R})$ as it would appear more natural. We will also see this is the only type of transformation in $GL(E^2)$ to require such a choice: the remaining, as will become apparent shortly, still have unit determinant.   

As an aside, we introduce how the transformations ${Q_V}$ act on a general element $t^{m}= {\left( \begin{smallmatrix} t^{\mu} & t \end{smallmatrix} \right)}^{tr}$, $\mu=1,\cdots,4$. It will prove worthy for the next chapter: 
\begin{align}
& t^{\mu'}  = (\det V)^{\frac{1}{2}}V_{\mu}^{\mu'} t^{\mu} \Rightarrow t^{\mu} \in (\det TM)^{\frac{1}{2}} \otimes TM \nonumber \\
& t' = (\det V)^{-\frac{1}{2}}t \Rightarrow t \in (\det TM)^{-\frac{1}{2}} \nonumber \\
\Rightarrow & t^m \in (\det TM)^\frac{1}{2} \left(TM \oplus (\det TM)^{-1}\right). \label{eq:t}
\end{align}
We would now like to reproduce the so-called \emph{shear-transformations in the} $T^{\ast}M$ \emph{direction} present in the $TM \oplus T^{\ast}M$ case (only this time it will be the $\wedge^{2}T^{\ast}M$ \emph{direction}). Let us introduce $e^{Q_{B}} \in GL(E^2)$: $(Q_{B})^{mn} \equiv \left( \begin{smallmatrix} 0 & B^{\mu} \\ 0 & 0 \end{smallmatrix} \right)$.
Then, consider $x^{m'n'}=(e^{Q_{B}})_{m}^{m'}(e^{Q_{B}})_{n}^{n'}x^{mn}$ given in matrix form by:
\begin{align} 
x^{m'n'}= x^{mn}+ 
\left( \begin{array}{cc} B^{\mu_1} X ^{\mu_2} - B^{\mu_2} X ^{\mu_1} & 0\\0&0 \end{array} \right). 
\end{align}
So, $Q_{B}$ performs the shear transformation desired since it changes $x$ by a dual of a two-form $B'$ with components ${B'}^{\mu_1 \mu_2} \equiv 2{B}^{[ \mu_1}X^{\mu_2 ]}$. There is no loss of generality then by regarding the form itself, call it $\tilde{B}$, as a map $\tilde{B} \colon TM \to \wedge^{2}T^{\ast}M$ and as a three-form in $M$ via $\tilde{B}(X)=i_{X} \tilde{B}$, in agreement with the treatment done for $TM \oplus  \wedge^p T^{\ast}M$.

A similar treatment can be done for \emph{shears in the} $\wedge^2 TM$ \emph{direction}. Let $e^{Q_{\beta}} \in GL(E^2)$:
$(Q_{\beta})^{mn} \equiv \left( \begin{smallmatrix} 0 & 0 \\ \beta_{\nu} & 0 \end{smallmatrix} \right)$. Again, taking the transformation $x'=e^{Q_{\beta}}x(e^{Q_{\beta}})^{tr}$ we obtain:
\begin{align} \label{eq:betatransf}
x^{m'n'}=x^{mn}+ \left( \begin{array}{cc} 0 & -\beta_{\nu}x^{\nu \mu_1} \\ \beta_{\nu}x^{\nu \mu_2} & 0 \end{array} \right). 
\end{align}
$\beta$ can be regarded as a map taking elements of $(\det TM) \otimes \wedge^2 T^{\ast}M \simeq \wedge^2 TM$ (or more generally $(2,0)$-tensors) to elements of $TM$ and there is a straightforward interpretation of the operation it conducts: $\beta(x)= i_{\beta}x$, with $i_{\beta} \colon \wedge^{p+1}TM \to \wedge^{p} TM$.    

We have seen how all the transformations studied for $O(E)$ seem also to be present in this case, although the decomposition is now given by $GL(5,\mathbb{R}) \simeq GL(4,\mathbb{R}) \oplus \exp (\wedge^3 T^{\ast}M) \oplus \exp (\wedge^3 TM)$ and the embedding of $GL(TM)$ emerged in a relatively different fashion. If $V \in GL^{+}(TM)$ we can still interpret it, in a sense, as an exponential term. For $V=e^{A}$, we can write:
\begin{align}
Q_V &= (\det V)^{\frac{1}{2}} \left( \begin{array}{cc} e^{A} & 0 \\ 0 & e^{-\mbox{tr }A} \end{array}\right) \nonumber \\
&= e^{\frac{1}{2} \mbox{tr A}}e^{Q_A}, 
\mbox{  }Q_A \equiv \left( \begin{array}{cc} A & 0 \\ 0 & -\mbox{tr }A \end{array}\right). 
\end{align}

\section{Differential structure} \label{sec:CoDerivative}
 
In this section we have as a goal to provide our space $E^2$ with a differential structure that should naturally be invariant under transformations in $GL(E^2)$. The essential step will be to introduce an \emph{affine connection} $\nabla$,
\begin{align}
\nabla \colon & E^2 \times E^2 \to E^2 \nonumber \\
& (x,y) \mapsto \nabla_x y,
\end{align}
that shall satisfy the properties (\cite{Nakahara}, $p.249$):
\begin{align*}
\nabla_x (y+z) &= \nabla_x y + \nabla_x z ,
\nabla_{x+y} z = \nabla_x y + \nabla_y z ,\\
\nabla_{fx} y  &= f (\nabla_x y) ,
\nabla_x (fy)  = x[f] + f (\nabla_x y),
\end{align*}
with $x,y,z \in E^2$, $f \in C^{\infty}(M)$ and the natural action of a generalized vector over a function being given by $x[f]=(X+\xi)[f]=X[f]+ f \wedge \xi$. We now proceed by investigating what those properties imply in a component form but first, let us introduce the differential operator $\partial_{mn} \in (E^2)^{\ast}$,
\begin{align} \label{def:gendiff}
\partial_{mn}\equiv \left(
\begin{array}{cc}
0 & -\partial_{\mu} \\
\partial_{\mu} & 0
\end{array} \right),
\end{align}
generalizing the usual one $\partial_{\mu} \in T^{\ast}M$ and antisymmetric in $m,n$. This choice certainly arises from consistency, given the symmetry of the generalized vectors themselves, but also from the study performed for the Courant bracket\footnote{See section \ref{sec:Courant} and Appendix \ref{Ap:D}}. Modifying our notation slightly, i.e., using an index $A=1,\cdots,10$ instead of $(m,n)$ and choosing as a basis for $E^2$ the set $\lbrace e_A \rbrace$, with $e_A$ represented by the matrix with zero in all the entries except for $1$ and $-1$ in the positions $(m,m+1)$ and $(m+1,m)$, respectively. We then define:
\begin{align}
\nabla_{e_A}e_B \equiv \nabla_A e_B \equiv \Omega_{AB}^C e_C.
\end{align}  
Using the defining properties above, one finds:
\begin{align}
\nabla_x y &= x^{A}(e_A[y^{B}]+\Omega_{AC}^B y^C)e_B \\
&= x^{A}(\partial_A y^{B}+\Omega_{AC}^B y^C)e_B, \nonumber
\end{align}
with the last equality accomplished by noting the form part of the combination $e_A[y^{B}]$ always vanishes. To end this study we come back to our original purpose: taking the differential structure to be invariant under $GL(5,\mathbb{R})$. This we can establish by defining an action of the connection components over $E^2$ has follows (now using the $mn$ indices):
\begin{align}
(\cdot) \colon & S_{\Omega} \times E^2 \to E^2 \nonumber \\
& (\Omega_{(pq)(rs)}^{mn},x^{rs}) \mapsto (\Omega \cdot x)_{pq}^{mn} \equiv 
(\Omega_{pq})^m_r x^{rn}-(\Omega_{pq})^n_r x^{rm},
\end{align}
with $S_{\Omega}$ designating the space for the connection components. 

The introduction of a covariant derivative follows trivially from the construction of the affine connection and in the case of generalized vectors they coincide. We note the action for the connection components defined above is in agreement with the action of $\nabla_x$ over $t \in (\det TM)^{\frac{1}{2}}\left( TM \oplus (\det TM)^{-1} \right)$:
\begin{align}
\nabla_x t = x^{pq}(\partial_{pq}t^m + (\Omega_{pq})^m_n t^n)e_m,
\end{align} 
with $\lbrace e_m \rbrace_{m=1,\cdots,5}$ as the canonical basis.

\section{Courant bracket} \label{sec:Courant}

We would like to be able to express our Courant bracket in terms of the general elements $x \in TM \oplus \wedge^{2}T^{\ast}M$ introduced in section \ref{sec:TMlinear}. To do that we will use our knowledge of the isomorphism\footnote{See Appendix \ref{Ap:A}} $\wedge^{2}T^{\ast}M \simeq (\det TM) \otimes \wedge^{2}T^{\ast}M$ to rewrite the definition \eqref{def:Courant}:
\begin{align}
\llbracket x^{5\mu} + \alpha x^{\mu_1 \mu_2},y^{5\mu}+ \alpha y^{\mu_1 \mu_2} \rrbracket 
= & {\lbrack x,y \rbrack}^{5\mu}  + {\left( {\alpha}\mathcal{L}_{x}y - {\alpha}\mathcal{L}_{y}x - \frac{\alpha}{2} d(i_x y - i_x y) \right)}^{\mu_1 \mu_2} \\
= & \left( x^{5\nu}\partial_{\nu}y^{5\mu} - y^{5\nu}\partial_{\nu}x^{5\mu} \right)
+ \alpha \left( x^{5\nu}\partial_{\nu}y^{\mu_1 \mu_2} - y^{5\nu}\partial_{\nu}x^{\mu_1 \mu_2} \right) \\
& - \alpha \frac{3}{2} \left( x^{[\mu_1 \mu_2}\partial_{\nu}y^{5\nu]} - y^{[\mu_1 \mu_2}\partial_{\nu}x^{5\nu]}
+ x^{[5\nu}\partial_{\nu}y^{\mu_1 \mu_2]} - y^{[\nu}\partial_{\nu}x^{\mu_1 \mu_2]} \right). \nonumber
\end{align}

The expression of the bracket in components allows us to write it as a general element $\llbracket x,y \rrbracket \in TM \oplus \wedge^{2}T^{\ast}M$ as follows:
\begin{align}
\llbracket x,y \rrbracket^{mn}& = \left( 
\begin{array}{cc}
\begin{array}{c}
\alpha (x^{5\nu}\partial_{\nu}y^{\mu_1 \mu_2} - y^{5\nu}\partial_{\nu}x^{\mu_1 \mu_2}) \\
+ \alpha \frac{3}{2} (x^{[\mu_1 \mu_2}\partial_{\nu}y^{\nu]5} - y^{[\mu_1 \mu_2}\partial_{\nu}x^{\nu]5}) \\
+ x^{5[\nu}\partial_{\nu}y^{\mu_1 \mu_2]} - y^{5[\nu}\partial_{\nu}x^{\mu_1 \mu_2]}
\end{array}
& - (x^{5\nu}\partial_{\nu}y^{5\mu_1} - y^{5\nu}\partial_{\nu}x^{5\mu_1}) \\
x^{5\nu}\partial_{\nu}y^{5\mu_2} - y^{5\nu}\partial_{\nu}x^{5\mu_2} & 0
\end{array} \right),
\end{align}
and equivalently as
\begin{align}
\llbracket x,y \rrbracket^{mn}& = \left( 
\begin{array}{cc}
\frac{\alpha}{2} \left( 
\begin{array}{c}
x^{5 \nu}\partial_{\nu}y^{\mu_1 \mu_2} - y^{5 \nu}\partial_{\nu}x^{\mu_1 \mu_2} \\ + \mbox{w.c.p.}(5,\nu,[\mu_1,\mu_2]) 
\end{array} 
\right)
& - \left( x^{5 \nu}\partial_{\nu}y^{5 \mu_1} - y^{5 \nu}\partial_{\nu}x^{5 \mu_1} \right) \\
x^{5 \nu}\partial_{\nu}y^{5 \mu_2} - y^{5 \nu}\partial_{\nu}x^{5 \mu_2} & 0
\end{array} \right), 
\end{align}
where we are using the shorthand w.c.p.$(5,\nu,[\mu_1,\mu_2])$ to denote the cyclic permutations of $(5,\nu,\mu_1,\mu_2)$ where the indices $\mu_1$, $\mu_{2}$ are anti-symmetrized and the remaining terms have weights of $2$, $-1$ and $2$, respectively.

Using the generalized differential operator presented in \eqref{def:gendiff} and the expression above we can present the bracket $\llbracket x,y \rrbracket^{mn}$ in a neat way\footnote{For details of the calculation see Appendix \ref{Ap:D}}:
\begin{align}
{\llbracket x,y \rrbracket}^{mn} 
& = \frac{1}{4}(x^{[m|p}\partial_{pq}^{A}y^{q|n]}-y^{[m|p}\partial_{pq}^{A}x^{q|n]})+\mbox{w.c.p.}(p,q,[n,m]).
\end{align}

Still following the coordinate approach we have developed till now, we will study the symmetries of the Courant bracket, in particular, the $B$-field transformations where $B$ is a smooth three-form. We would like to verify the result $e^{Q_B}\llbracket X+\xi,Y+\eta \rrbracket = \llbracket e^{Q_B}(X+\xi),e^{Q_B}(Y+\eta) \rrbracket$ iff $dB=0$ holds in the matrix format we have developed, as it should. It will prove to be to our advantage use from the beginning the dual representation for $B$ (more precisely noting that $\frac{1}{2}\epsilon^{\mu_1 \cdots \mu_4}x^{5\nu}B_{\nu \mu_3 \mu_4}=-2B^{[\mu_1}x^{\mu_2]5}$ where $B_{\nu \mu_3 \mu_4}=\epsilon_{\sigma \nu \mu_3 \mu_4}B^{\sigma}$). In what follows we use the definition ${x'}^{\mu_1 \mu_2} \equiv x^{\mu_1 \mu_2} + \frac{1}{2}\epsilon^{\mu_1 \cdots \mu_4}x^{5\nu} B_{\nu \mu_3 \mu_4}$:
\begin{align}
{\left\llbracket e^{Q_B}x(e^{Q_B})^{tr},e^{Q_B}y(e^{Q_B})^{tr}\right\rrbracket}^{mn} & = \left\llbracket
\left( \begin{array}{cc} {x'}^{\mu_1 \mu_2} & -x^{5\mu_1} \\ x^{5\mu_2} & 0 \end{array} \right),
\left( \begin{array}{cc} {y'}^{\mu_1 \mu_2} & -y^{5\mu_1} \\ y^{5\mu_2} & 0 \end{array} \right)
\right\rrbracket \\
& = {\llbracket x,y \rrbracket}^{mn} + 2 \left( \begin{array}{cc}
{\begin{array}{c}(x^{5\nu} \partial_{\nu} y^{5[\mu_2} - y^{5\nu} \partial_{\nu} x^{5[\mu_2})B^{\mu_1]} \\
- x^{5[\mu_1}y^{\mu_2]5}\partial_{\nu}B^{\nu} \end{array}} & 0 \\ 0 & 0
\end{array} \right). \nonumber
\end{align}
These last two terms need a more attentive look. Let us concentrate on the terms we would be expecting:
\begin{align}
i_x i_y d \colon & \wedge^{3}T^{\ast}M \to (\det M) \otimes \wedge^{2} T^{\ast}M \nonumber \\
& \frac{1}{3!} B^{\mu_1 \mu_2 \mu_3} \mapsto (i_x i_y dB)^{\nu_1 \nu_2} = x^{5[\nu_1} y^{\nu_2]5} \partial_{\nu}B^{\nu} 
\end{align}
\begin{align}
i_{[x,y]} B \colon & \wedge^{3}T^{\ast}M \to (\det M) \otimes \wedge^{2} T^{\ast}M \nonumber \\
& \frac{1}{3!} B^{\mu_1 \mu_2 \mu_3} \mapsto (i_{[x,y]} B)^{\nu_1 \nu_2} = 2(x^{5\nu} \partial_{\nu} y^{5[\mu_2} - y^{5\nu} \partial_{\nu} x^{5[\mu_2})B^{\mu_1]}.
\end{align}
This means, we in fact obtained:
\begin{align}
{\left\llbracket e^{Q_B}x(e^{Q_B})^{tr},e^{Q_B}y(e^{Q_B})^{tr}\right\rrbracket}^{mn} & ={\llbracket x,y \rrbracket }^{mn} + (i_{[x,y]} B)^{\mu_1 \mu_2} -2(i_x i_y dB)^{\mu_1 \mu_2} \\
& = {\left( e^{Q_B}\llbracket x,y \rrbracket (e^{Q_B})^{tr} \right)}^{mn} \mbox{ iff } dB = 0, \nonumber
\end{align}
which is to say we have accomplished the known result \eqref{eq:CourantB} of the coordinate free approach.  

\chapter{In search for a generalized metric} \label{ch:metric}

In analogy with the case of the generalized bundle $E$ we would like to introduce a metric $G$ for $E^2$ generalizing the concept we have for the usual bundle $TM$. However, there is not a canonical way of introducing this structure as there was for $TM \oplus T^{\ast}M$. To make our choice we will rely in two basic features: we want this new object to still transform under $GL(5,\mathbb{R})$ and we want it to include a usual metric $g \in S^{2}T^{\ast}M$, as it happened for the $E$ case. 

Since we are interested in describing supergravity solutions, it makes sense to consider such $g$ with a Lorentzian signature. We can then have two choices for the overall signature: $(1,4)$ or $(2,3)$. We opt by the latter motivated by two factors. The first is the natural embedding of $AdS_4$, one of our three-form supergravity solutions, in $\mathbb{R}^{2,3}$. The second relies on the need for spinors, representations of a Clifford algebra\footnote{See appendices \ref{Ap:B} and \ref{Ap:C}} that is in this case $\mathpzc{Cl}(1,3)$, to describe supergravity. This algebra includes $Spin(2,3)$ as a subgroup\footnote{See Appendix \ref{Ap:C}} and this again brings us to a $(2,3)$ signature, since such a representation might be of importance when considering a full $4d$-supergravity action, instead of the merely bosonic one.

This chapter can be viewed as devoted to two main purposes: a more technical one that corresponds to a treatment similar to the one undertaken in sections \ref{sec:GenMetricT1} and \ref{sec:TMlinear} where we also include the description of generalized Killing vector fields and the study of the inner products $G$ will give rise to; and a more practical, where we apply the formalism to the specific case of three-form supergravity, comprising the last three sections.  

\section{Linear algebra}

We would like to accommodate a term involving a (inverse of a) metric, $g'^{\mu_1 \mu_2}$, that may not be $g$, the one we suppose our Lorentzian manifold $M$ is endowed with; a vector field $A^{\mu}$ and a scalar field $\varphi$ on a general symmetric object $G^{mn}$. We start under the assumption $G^{-1} \in TM \oplus S^{2}TM \oplus (\det TM)$. Since we would like to treat $x^{mn}$ and $G^{mn}$ as the same ``type" of object, having the same kind of transformation becomes a requisite. Let us say we have,
\begin{align}
G^{mn}= \left( \begin{array}{cc} g'^{\mu_1 \mu_2} & A^{\mu_1} \\ A^{\mu_2} & \varphi \end{array} \right),
\end{align}
we now consider the same transformation $Q_V \in GL (E^2)$ we did before and begin the process once again. If we start with the transformation for $A^{\mu}$, we get $G^{5 \mu'}=V_{\mu}^{\mu'}G^{5 \mu}$, so $A^{\mu}$ already transforms appropriately: $G^{5 \mu} \in TM$. In the case of the scalar we obtain $G'^{55}=(\det V)^{-1}G^{55}$($\Rightarrow G^{55} \in$ ($\det TM)^{-1}$). This means we should accommodate a factor of $|\tilde{g}|^{-\frac{1}{2}} \varphi$ instead of $\varphi$, remembering that the determinant of the metric, $\tilde{g}$, scales with the square of the determinant of the linear combination. To determine the transformation laws completely, we should check the metric transforms appropriately: $G^{\mu_1'\mu_2'}= (\det V)V_{\mu_1}^{\mu_1'}V_{\mu_2}^{\mu_2'}G^{\mu_1 \mu_2}$ implying $G^{\mu_1 \mu_2 } \in (\det TM) \otimes S^{2}TM$. This means $G^{\mu_1 \mu_2}$ should be set as a density of weight $+1$: we see that $G^{\mu_1 \mu_2}=|\tilde{g}|^{\frac{1}{2}}g'^{\mu_1 \mu_2}$ is the correct combination. 

To summarize, we have discovered the appropriate object $G^{mn}$ should be:
\begin{align}
G^{mn} = |\tilde{g}|^{\frac{1}{2}} \varphi \left( \begin{array}{cc}
g'^{\mu_1 \mu_2} & |\tilde{g}|^{-\frac{1}{2}}A^{\mu_1} \\ |\tilde{g}|^{-\frac{1}{2}}A^{\mu_2} & |\tilde{g}|^{-1}
\end{array} \right), 
\end{align}
i.e., belong to $(\det TM)\otimes ((\det TM)^{-1}TM \oplus S^{2}TM \oplus (\det TM)^{-2})$. Note we have redefined ${g'}^{\mu_1 \mu_2}$ and $A^{\mu}$ to include a factor of ${\varphi}^{-1}$. 

We would now like to express the generalized (inverse) metric, $G^{mn}$, in a way such that $(G_B)^{-1}=e^{Q_{B}}G^{-1}_{0}(e^{Q_{B}})^{tr}$, for a given $B$-field where $G^{-1}_{0}$ is defined to be:
\begin{align}
(G_{0})^{mn} \equiv |\tilde{g}|^{\frac{1}{2}} \varphi \left( \begin{array}{cc}
g^{\mu_1 \mu_2} & 0 \\ 0 & |\tilde{g}|^{-1}
\end{array} \right).
\end{align}
Note how the choice of $|\tilde{g}|^{-\frac{1}{2}} \varphi$ as the $G^{55}$ element brings us to a $(2,3)$ signature in accordance with our motivation. Also, now we are indeed using the usual metric $g$. The study we made for $x^{mn}$ motivates $e^{Q_{B}}$ to be a shear transformation in the $\wedge^{2}T^{\ast}M$ direction, since the relevant object for us is a three-form. Under the assumption $B^{\mu} \in (\det TM) \otimes T^{\ast}M$:
\begin{align}
(G_B)^{mn} \equiv {\left(e^{Q_{B}}G^{-1}_{0}(e^{Q_{B}})^{tr}\right)}^{mn}=|\tilde{g}|^{\frac{1}{2}}\varphi \left( {\begin{array}{cc}
g^{\mu_1 \mu_2} + |\tilde{g}|^{-1}B^{\mu_1}B^{\mu_2} &|\tilde{g}|^{-1}B^{\mu_1} \\
|\tilde{g}|^{-1}B^{\mu_2} & |\tilde{g}|^{-1} 
\end{array}}\right),
\end{align}
all the components of $(G_B)^{mn}$ transform according to what was expected and we find the relations $g'^{\mu_1 \mu_2}=g^{\mu_1 \mu_2} + |\tilde{g}|^{-1}B^{\mu_1}B^{\mu_2} $ and $B^{\mu}=|\tilde{g}|^{\frac{1}{2}}A^{\mu}$. Since the condition $(G_B)^{mn}(G_B)_{nl}=\delta_{l}^{m}$ must be satisfied, parting from:
\begin{align}
(G_0)_{mn} = |\tilde{g}|^{-\frac{1}{2}} \varphi^{-1} \left( 
\begin{array}{cc}
g_{\mu_1 \mu_2} & 0 \\ 0 & |\tilde{g}|
\end{array} \right),
\end{align}
we obtain:
\begin{align}
(G_B)_{mn} 
& = \left( e^{Q_{B}}G^{-1}_{0}(e^{Q_{B}})^{tr}\right)^{-1}_{mn} 
= \left( \left(e^{-Q_{B}})^{tr}G_{0}(e^{-Q_{B}} \right) \right)_{mn} \\
& = |\tilde{g}|^{-\frac{1}{2}} \varphi^{-1} \left( \begin{array}{cc}
g_{\mu_1 \mu_2} & - g_{\mu_1 \nu}B^{\nu} \\ - g_{\mu_2 \nu}B^{\nu} & |\tilde{g}| + B^{\mu}g_{\mu \nu}B^{\nu}
\end{array} \right). 
\end{align}
This brings us to the conclusion $G_{\mu_1 \mu_2} \in (\det TM)^{-1} \otimes S^{2}T^{\ast}M$, but what about the other transformations? We shall analyse the contraction $g_{\mu \nu}B^{\nu}$ in more detail. We know $B^{\nu} = \frac{1}{3!} \epsilon^{\nu \mu_1 \mu_2 \mu_3}B_{\mu_1 \mu_2 \mu_3}$. The contraction of interest is then:
\begin{align}
& g_{\mu \nu}B^{\nu}= \frac{1}{3!} \epsilon_{\mu}^{\mu_1 \mu_2 \mu_3} B_{\mu_1 \mu_2 \mu_3} = |\tilde{g}|^{\frac{1}{2}} (\star B)_{\mu} \\
\Rightarrow \mbox{ }& g_{\mu \nu}B^{\nu} \in (\det TM) \otimes T^{\ast}M. \nonumber
\end{align}
This means $G_{5 \mu} \in T{^\ast}M$ and $G^{55} \in (\det TM)$, meaning $G \in (\det TM)^{-1} \otimes ((\det TM)T^{\ast}M \oplus S^2 T^{\ast}M \oplus (\det TM)^2)$ as expected.  

For completeness we should include the shear transformations in the $TM$ direction $e^{Q_{\beta}}$, baring in mind that $\beta: T^{\ast}M \otimes T^{\ast}M \to TM$. The transformation $G_{\beta}^{-1}=e^{Q_{\beta}}G_{0}^{-1}(e^{Q_{\beta}})^{tr}$ holds:
\begin{align}
(G_{\beta})^{mn} & = |\tilde{g}|^{\frac{1}{2}} \varphi \left( {\begin{array}{cc}
g^{\mu_1 \mu_2} & g^{\mu_1 \nu} \beta_{\nu} \\ g^{\mu_2 \nu} \beta_{\nu} & g^{\mu \nu} \beta_{\mu} \beta_{\nu} + |\tilde{g}|^{-1}
\end{array}} \right) \\
(G_{\beta})_{mn} & = |\tilde{g}|^{-\frac{1}{2}} \varphi^{-1} \left( {\begin{array}{cc}
g_{\mu_1 \mu_2} + |\tilde{g}|\beta_{\mu_1} \beta{\mu_2} & - |\tilde{g}|\beta_{\mu_1} \\
- |\tilde{g}|\beta_{\mu_2} & |\tilde{g}|
\end{array}} \right).
\end{align}
According to the laws of transformation one would expect, we should have $\beta \in (\det TM)^{-1} T^{\ast}M \simeq \wedge^3 TM$ in accordance with the transformation \eqref{eq:betatransf}.  
 
\section{Generalized Killing vector fields}   

Following the same approach developed in section \ref{sec:genK} we can introduce the concept of generalized Killing vector fields for $TM \oplus \wedge^2 T^{\ast}M$. The only difference between the presentation for the cases of $E$ and $E^2$ is that in the latter we opt for a coordinate representation, once again using the isomorphism studied in Appendix \ref{Ap:A}. The expression of the generalized Lie derivative for the metric is given by:
\begin{align}
(\mathbb{L}_v G)_{mn} &=\left( \begin{array}{cc}
(L_{v}g)_{\mu_1 \mu_2} & - \left[ g(L_{v}B + 2dv)+(L_{v}g)B \right]_{5\mu_1} \\
- \left[ (L_{v}B+2dv)g+B(L_{v}g) \right]_{5\mu_2}
& \begin{array}{c}
 -L_{v}|\tilde{g}|-(L_{v}B+2dv)gB-B(L_{v}g)B \\ + Bg(2dv+L_{v}B) \end{array}
\end{array} \right)
\end{align} 
so, once more, if we take $v \to 2v$ one accomplishes the conditions $(\mathcal{L}_v B + dv)^{\mu}=0$ and $(\mathcal{L}_v g)_{\mu_1 \mu_2}=0 \Rightarrow \mathcal{L}_v |\tilde{g}| =0$ can be summarized by setting $\mathbb{L}_v G=0$. This is the definition, already presented in \eqref{eq:Killing}, of a generalized Killing vector field.

\section{Inner products}

When considering elements $t^m \in (\det TM)^\frac{1}{2} \left(TM \oplus (\det TM)^{-1}\right)$ there is a notion of a norm that can be introduced using the metric $G_0$. Setting $t^{m}=(t^{\mu},t^{5})=(|\tilde{g}|^{\frac{1}{4}}V^{\mu},|\tilde{g}|^{-\frac{1}{4}}V^5)$ such that $(V^{\mu},V^5)\in TM \oplus (\det TM)^{0}$, we obtain:
\begin{align}
(G_0)_{mn}t^{m}t^{n} &= \varphi^{-1} (V^2 + (V^5)^2). \label{eq:t_inner}
\end{align}
Note how the signature is what we would expect: $(2,3)$. For elements $x^{mn} \in TM \oplus \wedge^{2} T^{\ast}M$ a similar relation can be found:
\begin{align}
(G_0)_{mp}(G_0)_{qn}x^{mn}x^{pq} & = -\mbox{tr}(xG_0xG_0) \nonumber \\
& = |\tilde{g}|^{-1}\varphi^{-2}\left( x^{\mu_1 \mu_2}x_{\mu_1 \mu_2} + 2|\tilde{g}|x^{5\mu}x_{5\mu}\right) \nonumber \\
& = \varphi^{-2}\left(\xi^2 + 2X^2\right), \label{eq:Xinner} 
\end{align} 
for $x=X+\xi$. Again, the right type of signature is reflected. 

\section{Equivalence class of metrics and Parallelizable manifolds} \label{sec:Parallel}

It is our objective to be able to write supergravity solutions in the formalism of generalized geometry, so we shall bare in mind what we studied in chapter \ref{ch:membranes}. 

First, we note an ``initial" generalized metric $(G_{0})_{mn}$ should be composed of an ordinary metric and of a relevant scalar. A first candidate one could think of in this context for this last one would be the cosmological constant but it will already be encoded by the three-form $C^{\mu_1 \mu_2 \mu_3}$ introduced in \eqref{eq:Cform}. To make this more apparent we should first study the relation between $C_{\mu_1 \mu_2 \mu_3}$ and $\lambda$. From the definition $C^{\mu_1}= \frac{1}{3!} \epsilon^{\mu_1 \cdots \mu_4} C_{\mu_2 \mu_3 \mu_4}$ and using the relation $F_{\mu_1 \cdots \mu_4} = \lambda \varepsilon_{\mu_1 \cdots \mu_4} = 4 \partial_{\mu_1} C_{\mu_2 \mu_3 \mu_4}$ one arrives at $\lambda = |\tilde{g}|^{\frac{1}{2}} \partial_{\mu}C^{\mu}$. This means the information given by $\lambda$ can already be included, just by performing the correct $B$-field transformation (i.e. using $B^{\mu} \equiv C^{\mu}$). 

We can conjecture the elimination of one degree of freedom and such an idea is reinforced by a different perspective for the generalized metric of $E_{4(4)}$, presented by Hull (\cite{Hull}, $p.12$). Here, such an object appears parametrizing the coset space $SL(5,\mathbb{R})/SO(5)$ whereas on the approach we have developed $G \in GL(5,\mathbb{R})/SO(2,3)$. The two spaces have representations $\mathbf{10+4}$ and $\mathbf{10+4+1}$, respectively, and this strongly suggests the disappearance of the scalar liberty. To achieve this objective we opt to define an equivalence class of metrics. The equivalence relation is given by: 
\begin{align}
\left[ (G_{0})_{mn} \right] =
\left\lbrace 
(G^{eq}_{0})_{mn} \colon g^{eq}_{\mu_1 \mu_2} = \varphi_1 g_{\mu_1 \mu_2},\mbox{ } {\tilde{g}}^{eq}= \varphi_2 \tilde{g},\mbox{ } \forall_{\varphi_i \in {(\det TM)^0}_{+}, i=1,2} 
\right\rbrace ,
\end{align}
where $(\det TM)^0_{+}$ stands for the space of positively definite scalars (otherwise, the signature of our metric could be compromised). This brings $G$ to the space $GL(5,\mathbb{R})/ \left( SO(2,3) \otimes \mathbb{R}^{+} \right)$ and, we note, it is always possible to opt for a representative with unit determinant.

The non-coordinate (local) basis of vielbein we can usually find for a metric $g$ along with the inner product presented in \eqref{eq:t_inner} motivates the introduction of a similar notion for generalized metrics. To accomplish this objective let us first recall the notion of \emph{parallelizability}.

A manifold $M$ is called \emph{parallelizable} if there are $C^{\infty}$ vector fields $X_{1},\cdots,X_{d}$ defined on all of $M$ such that for every $q \in M$, ${X_{1}\mid_q,\cdots,X_{d}\mid_q}$ is a basis of $T_{q}M$ (\cite{Bishop}, $p.160$). The vector fields $\lbrace X_{i} \rbrace_{i=1,\cdots,m}$ are then called a \emph{parallelization} of $M$. In other words, a parallelizable manifold has a \emph{globally defined} frame. This feature is present in flat spaces $\mathbb{R}^{p,q}$ and we would like, in a sense, to bring it to solutions of supergravity actions, using generalized geometry. To do that we introduce a different notion. We say $M$ is $p$\emph{-parallelizable} if there is a globally defined basis of sections ${x_{1},\cdots,x_{d}}$ for $TM \oplus \wedge^p T^{\ast}M$. We shall note that upon the isomorphism $TM \simeq T^{\ast}M$ it is irrelevant if we choose to define \emph{tetrads} in terms of vector fields and forms, one-forms and multi-vectors or any other valid combination of duals. In particular, we will most commonly find a generalized frame in $T^{\ast}M \oplus \wedge^d T^{\ast}M$. 

For our case of relevance, i.e., $d=4$, we will represent an element of the basis, $E^{a}$, as a $5$-dimensional vector with the components of a one-form and a volume form. The matrix with these lines provides a representation for the generalized parallelization $E$.  

A representative of our equivalence class, say $G_0$, will always be able to be written as $G_0 \equiv F^{tr}\eta F=|\tilde{g}|^{-\frac{1}{2}} E^{tr} \eta E$ where $\eta=\mbox{diag}(1,-1,-1,-1,1)$ and $E$ is a global $4$-frame (or, upon Hodge duality, $0$-frame). Note how the equivalence relation has been set up in a way that makes sure all the generalized metrics on a class still define a $4$-parallelization.    

Suppose then we have found a global frame $E$ satisfying $G_0 =|\tilde{g}|^{-\frac{1}{2}} E^{tr} \eta E$. Given the relation $G_B=(e^{-Q_B})^{tr}G_0 e^{-Q_B}$ we see $Ee^{-Q_B}$ provides a generalized parallelization for $G_B$. This means if we can find a representative with nowhere vanishing tetrads, not only every element of the equivalence class will share this property but also every $B$-transform of the initial metric. We note in the case of $TM \oplus T^{\ast}M$ this construction is not so straightforward (see, for example, \cite{Grana}). This analysis extends naturally to $G_{\beta}$ and the inverse metric $G^{-1}$. 

\section{Generalized solutions of three-form supergravity} \label{sec:GSusy}

The ideas presented above will now be applied to our three-form supergravity solutions. We start by studying the case $\lambda^{2}=0$, i.e., the flat space scenario. We know the metric solution is $g_{\mu \nu}=\eta_{\mu \nu}$. It is then trivial to see that setting the representative $G_0$ to be $\eta=diag(1,-1,-1,-1,1)$ - the same representation the flat metric of $\mathbb{R}^{2,3}$ has in Cartesian coordinates $\lbrace x^m \rbrace_{m=1,\cdots ,5}$ - satisfies our requirements. The basis of generalized tetrads is 
\begin{align} \label{eq:flatGF}
\left\lbrace
E^{a}
\right\rbrace_{a=1,\cdots,5}
&=\left\lbrace 
E^{\mu} = dx^{\mu}, E^5 = dx^1 \wedge \cdots \wedge dx^4 
\right\rbrace_{\mu=1,\cdots,4},
\end{align}
where we have opted for the $4$-parallelization but we could equally well take the $0$-frame by considering the (Hodge dual) smooth function $1$ instead of the volume form $dx^1 \wedge \cdots \wedge dx^4$.
  
If $\lambda^{2} \neq 0$ we will consider the maximally symmetric solution for vacuum Einstein's equations with a negative cosmological constant - \emph{Anti de Sitter} space. 
We can define $AdS_4$ as the quadric surface $x \eta x^{tr}+\frac{3}{\lambda^2}=0$ embedded in ${\mathbb{R}}^{2,3}$. Let us present a diffeomorphism expressing the Cartesian coordinates  for ${\mathbb{R}}^{2,3}$ in terms of the set $\lbrace y^{m}(z^{\mu}),R \rbrace_{m=1,\cdots,5}$ where $z^{\mu},\mu=1,\cdots,4$ are coordinates for $AdS_4$:
\begin{align}
f \colon & M \to M \nonumber \\
& x^m \mapsto f(x^m)=\frac{\lambda^2}{3}R y^m,
\end{align}
with the conditions $\eta_{mn} y^m y^n=\frac{3}{\lambda^2}$, $m,n=1,\cdots,5$ verified. It brings the (flat) metric to the (at least block diagonal) form:
\begin{align}
f_{\ast}g^{{\mathbb{R}}^{2,3}} 
& = \frac{\lambda^2}{3}Rg^{AdS_4}+ dR \otimes dR. \nonumber
\end{align}
This metric has singularities (for example, consider $R \to 0$) but these are just coordinate singularities: there is a global frame for this space. It emerges explicitly by taking the Cartesian coordinates and searching for a basis of forms globally defined over the cotangent bundle. A trivial example is then $\lbrace e^{a}=dx^{a} \rbrace_{a=1,\cdots,5}$. The key idea for constructing a $4$-parallelization for the quadric surface (to which only local frames can be defined) is to take advantage of this feature of the ambient space. 

The form fields of $\mathbb{R}^{2,3}$ are still well defined when restricted to $AdS_4$, even though when pulled-back with the diffeomorphism presented they may present some (coordinate) singularities:
\begin{align}
f_{\ast}dx^m & = \frac{\lambda^2}{3}(Rdy^m + y^m dR).
\end{align}
This is still a global frame in the new coordinate system, although we have lost the nice diagonal representation. 

Now, to set a generalized global frame we shall let its elements $E^{a}$ have a similar structure to the one presented, that we know is well defined, by considering $vol_g \equiv |{\tilde{g}}^{AdS_4}|^{\frac{1}{2}} dz^1 \wedge \cdots \wedge dz^4$, the metric volume form of $AdS_4$, to play the same role the radial coordinate $R$ had in $\mathbb{R}^{2,3}$. We must try and make the process more apparent: let $J_f = \left( \begin{smallmatrix} R \frac{\partial y^m}{\partial z^{\mu}} & y^m \end{smallmatrix} \right)$ be the Jacobian matrix of the coordinate transformation introduced by $f$. It gives us a representation of a global frame for $\mathbb{R}^{2,3}$ and we shall ``borrow" it to represent our generalized global frame as follows:
\begin{align}
e &= \left( \begin{array}{cc} R \frac{\partial y^m}{\partial z^{\mu}} & y^m \end{array} \right)
\left( \begin{array}{c} dz^{\mu} \\ dR \end{array} \right) 
\to E = \left( \begin{array}{cc} \frac{\partial y^m}{\partial z^{\mu}}  & y^m |\tilde{g}^{AdS_4}|^\frac{1}{2} \end{array} \right)
\left( \begin{array}{c} dz^{\mu} \\ dz^{1} \wedge \cdots \wedge dz^{4} \end{array} \right).
\end{align}

In other words, we are making sure there is a generalized global frame by giving the space it lives in the structure of one - for instance $T^{\ast}{\mathbb{R}}^{2,3}$ - where we know it exists. 
    
With this information in mind, let us come back to a possible representative for the generalized metric:
\begin{align}
(G_0)_{mn} = |{\tilde{g}}^{AdS_4}|^{-\frac{1}{2}} \left( \begin{array}{cc} g^{AdS_4}_{\mu_1 \mu_2} & 0 \\ 0 & |{\tilde{g}}^{AdS_4}| \end{array} \right).
\end{align}
 
Let $\varphi_1$, $\varphi_2$ be scalars in $(\det TM)^0_+$. The problem now reduces to find a generalized parallelization for $AdS_4$ satisfying $E^{a}=\varphi_1 y^a vol_g + \varphi_2 dy^a$. We shall make this construction explicitly, taking coordinates $\lbrace z^{\mu} \rbrace_{\mu=1,\cdots,4}= \lbrace \tau, \rho, \theta, \phi \rbrace$. In this case the metric reads\footnote{For the explicit coordinate transformation see Appendix \ref{Ap:E}.}:
\begin{align}
g^{AdS_4}=\frac{3}{\lambda^2} \left(
\cosh^2 \rho d\tau \otimes d\tau
-d\rho \otimes d\rho
-\sinh^2 \rho d\theta \otimes d\theta
-\sinh^2 \rho \sin^2 \theta d\phi \otimes d\phi
\right),
\end{align}
and from the pull-back of the global frame $\lbrace e^{a} \rbrace_{a=1,\cdots,5}$ we obtain the conditions $\varphi_1^2=\varphi_2^2=1$ for this particular representative. To sum up, we can choose as our global frame the set:
\begin{align}
\lbrace E^1 &=- \sin \tau \cosh \rho d\tau + \cos \tau \sinh \rho d\rho + \cos \tau \cosh \rho vol_g,\nonumber \\
E^2 &= \cosh \rho \sin \theta \cos \phi d\rho + \sinh \rho \cos \theta \cos \phi d\theta - \sinh \rho \sin \theta \sin \phi d\phi + \sinh \rho \sin \theta \cos \phi vol_g , \nonumber \\
E^3 &= \cosh \rho \sin \theta \sin \phi d\rho - \sinh \rho \cos \theta \sin \phi d\theta + \sinh \rho \sin \theta \cos \phi d\phi + \sinh \rho \sin \theta \sin \phi vol_g , \nonumber \\
E^4 &= \cosh \rho \cos \theta d\rho - \sinh \rho \sin \theta d\theta + \sinh \rho \cos \theta vol_g, \nonumber \\
E^5 &= \cos\tau \cosh\rho d\tau + \sin \tau \sinh \rho d\rho + \sin \tau \cosh \rho vol_g \rbrace.
\end{align}
Given a generalized parallelization for $G_0$ it is straightforward to construct a basis $\lbrace \tilde{E}_a \rbrace$ for $G_0^{-1}=|\tilde{g}|^{\frac{1}{2}} \tilde{E}^{tr} \eta \tilde{E}$ since $E \tilde{E}^{tr}=\mathds{1} \Rightarrow G_0G_0^{-1}=E^{tr} \eta E \tilde{E}^{tr}\eta \tilde{E}=\mathds{1}$. From what we have studied before this is equivalent to define $\tilde{E}_a$ using $\tilde{e}_a=f^{\ast}\frac{\partial}{\partial x^{a}}$, the basis of globally defined vector fields for $f({\mathbb{R}}^{2,3})$ in a way identical to the one introduced for $G_0$.
  
In both cases we were able to assemble a global $4$-frame in the context of finding a basis for the generalized metric but we can go further. The next section is devoted to another project: the construction of a $2$-parallelization, remarkably providing us with a nowhere vanishing basis for $E^2$.

\section{Generalized global frames in the construction of a basis for $E^2$} \label{sec:genK2}

Thanks to the developments presented in \ref{sec:GSusy}, we now have generalized global frames to the maximally symmetric solutions of our toy model supergravity. Recall one of the features of the notion is being independent of the existence of a metric even though its introduction was motivated by the generalized one. 

What we propose ourselves to do in this final section is to construct a basis for the generalized tangent bundle, hopefully again in a canonical way, using our knowledge of the $4$-parallelizations. Flat space will be the trivial example, followed by a more meaningful discussion: $AdS_4$. 

Remembering the definition of $F^{a}=|\tilde{g}^{AdS_4}|^{-\frac{1}{4}}E^{a}$ as satisfying the relation:
\begin{align}
G_{mn}&= \eta_{ab}F^{a}_m F^{b}_n,
\end{align}
and considering the inner product defined in \eqref{eq:Xinner} we can construct the analogue of this expression for elements $F^{ab} \in (E^2)^{\ast}$:
\begin{align} \label{eq:Xmetric}
G_{mn}G_{pq}-G_{mq}G_{pn}=(\eta_{ac}\eta_{bd}-\eta_{ad}\eta_{bc})(F^{ab})_{mp}(F^{cd})_{nq}.
\end{align}
It is easy to verify that setting $(F^{ab})_{mn} \equiv F^{[a}_{m} F^{b]}_{n}=|\tilde{g}^{AdS_4}|^{-\frac{1}{2}}E^{[a}_{m} E^{b]}_{n}$ will satisfy the above expression. Also follows $\tilde{F}_{ab} \in E^2$ satisfies a relation analogous to \eqref{eq:Xmetric} and is decomposable as $(\tilde{F}_{ab})^{mn}=\tilde{F}_{[a}^m \tilde{F}_{b]}^n$. The two basis satisfy the relations:
\begin{align}
(\tilde{F}^{ab})_{mp}(F_{cd})^{pm}&=\delta_{[d}^{a}\delta_{c]}^{b} \label{eq:F1}\\
(F_{ab})^{mp}(\tilde{F}^{ba})_{rs}&=\delta_{[s}^{m}\delta_{r]}^{p} \label{eq:F2},
\end{align}
as one can check directly using the properties of the global frames.

We know recover the generalized global frame (taking the $0$-parallelization) for flat space, $\mathbb{R}^{1,3}$ given in equation \eqref{eq:flatGF}: 
\begin{align}
\left\lbrace E^{a} \right\rbrace_{a=1,\cdots,5} & = \left\lbrace E^{\mu}=dx^{\mu},E^5=1 \right\rbrace_{\mu=1,\cdots,4}, \\
\left\lbrace \tilde{E}_{a} \right\rbrace_{a=1,\cdots,5} &= \left\lbrace \tilde{E}_{\mu}=\frac{\partial}{\partial x^{\mu}}, \tilde{E}_{5}=1 \right\rbrace_{\mu=1,\cdots,4}.
\end{align}
Is trivial to identify both $F^{A} \equiv F^{ab} \in (E^2)^{\ast}$ and $\tilde{F}_{A} \equiv \tilde{F}_{ab} \in E^2$:
\begin{align}
\left\lbrace F^{A} \right\rbrace_{A=1,\cdots,10} & =
\left\lbrace
F^{\mu_1 \mu_2}=\left (\begin{array}{cc}
dx^{\mu_1} \wedge dx^{\mu_2} & 0 \\
0 & 0
\end{array} \right),
F^{5\mu }=\left( \begin{array}{cc}
0 & dx^{\mu} \\
-dx^{\mu} & 0
\end{array} \right) \right\rbrace_{\mu,\mu_i=1,\cdots,4}, \\
\left\lbrace \tilde{F}_{A} \right\rbrace_{A=1,\cdots,10} &=
\left\lbrace
\tilde{F}_{\mu_1 \mu_2}=\left (\begin{array}{cc}
\epsilon_{\mu_1 \cdots \mu_4} dx^{\mu_3} \wedge dx^{\mu_4} & 0 \\
0 & 0
\end{array} \right),
F^{5\mu }=\left( \begin{array}{cc}
0 & \frac{\partial}{\partial x^{\mu}} \\
-\frac{\partial}{\partial x^{\mu}} & 0
\end{array} \right) \right\rbrace_{\mu,\mu_i=1,\cdots,4}.
\end{align}

For the case of $AdS_4$ a general element of the $0$-parallelization reads 
\begin{align}
E^{a}& = \left (\begin{array}{c}
\frac{\partial y^a}{\partial z^{\mu}} dz^{\mu}\\ y^{a}|\tilde{g}^{AdS_4}|^{\frac{1}{2}}
\end{array} \right),
\end{align}
using the conventions of the preceding section for the coordinates. We arrive at the representation of $F^{ab} \in (E^2)^{\ast}$ in a straightforward way,
\begin{align} \label{eq:E_2_star_basis}
(F^{ab})_{mn}= \left( \begin{array}{cc}
|\tilde{g}^{AdS_4}|^{-\frac{1}{2}}\partial_{\mu_1} y^{[a} \partial_{\mu_2} y^{b]}&
-y^{[a} \partial_{\mu_1} y^{b]} \\
y^{[a} \partial_{\mu_2} y^{b]}& 0
\end{array} \right),
\end{align}
remembering $\wedge^2 TM \simeq (\det TM)^{-1}\otimes \wedge^2 T^{\ast}M$, but exactly as happened for the global frames, $\tilde{F}_{ab}$ is harder to determine without an explicit set of coordinates $z^{\mu}$, although given those we can construct it rather easily using the relations \eqref{eq:F1} and \eqref{eq:F2}.   

\chapter{Final remarks}

This short last chapter tries to provide an account of the ideas worth developing following this dissertation, as well as offering additional insight over the ones already studied. We do not mean to state such ideas as original (most of these developments can be found under study in the literature, see \cite{GualtieriPhD} \cite{Grana}) but only to ascertain them as the natural path to pursue. In contrast with the overall technical nature of the work presented, this conclusion, mostly speculative, will not at all try to be rigorous in mathematical terms, and will mostly focus on a reanalysis of the last sections of chapter \ref{ch:metric}.

On chapter \ref{ch:1} we have introduced the Courant bracket \eqref{def:Courant} over sections as a generalization of the usual Lie bracket on vector fields. Given the structure of a Lie group that comes associated to the latter a question naturally arises: can we find the equivalent notion for the Courant bracket? Specifically, we would like to understand if certain manifolds of interest that are not Lie groups could be classified as ``Courant groups" in the sense we can find structure constants $c^C_{AB} $ satisfying $\llbracket x_A , x_B \rrbracket = c^C_{AB} x_C$, with $\lbrace x_A\rbrace$ a basis for $E^p$. It is a different issue to understand the relevance of $p$ in that definition, although our interest, in the context of a language for supergravity, would be in $p=1,2$. On a more concrete note, given the toy model we studied, it is certainly worth the effort to compute the Courant bracket for the $E^2$ basis of $\mathbb{R}^{1,3}$ and $AdS_4$. A similar analysis might be of interest for an extension of the Courant bracket, the \emph{twisted Courant Bracket} (\cite{GualtieriPhD}, $p.43$) which has the same symmetries:
\begin{align}
\llbracket X+\xi, Y + \eta \rrbracket_H & \equiv \llbracket X + \xi, Y + \eta\rrbracket + i_Y i_X H,
\mbox{  } H \in \wedge^{p+2}T^{\ast}M.
\end{align} 
Moreover, upon investigating such a structure it would benefit us find the analogous of the duality characteristic of (Lie group) vectors and one-forms, namely:
\begin{align}
[e_{\mu},e_{\nu}]&=c^{\sigma}_{\mu \nu} e_{\sigma} \leftrightarrow
d\tilde{e}^{\sigma} + \frac{1}{2} c^{\sigma}_{\mu \nu} \tilde{e}^{\mu} \wedge \tilde{e}^{\nu} =0,
\end{align}
where $\lbrace e_\mu \rbrace_{\mu=1,\cdots,d}$ and $\lbrace \tilde{e}^{\mu} \rbrace_{\mu=1,\cdots,d}$ are basis for $TM$ and $T^{\ast}M$. This relation would be of value, for example, to find $c^C_{AB}$ for the $AdS_4$ case we have studied, remembering how much easier it was to find a basis \eqref{eq:E_2_star_basis} for $(E^2)^{\ast}$.

Changing the matter at hand, we now come back to the introduction of $X^{mn}$ and $G^{mn}$ as antisymmetric and symmetric objects with the same group of transformations. This brings the idea of \emph{generalized tensors} in to form. Reinforcing it is the element $t^m$ presented in \eqref{eq:t} that is naturally acted upon with the covariant derivative of section \ref{sec:CoDerivative}. A need for a more detailed study of on how a complete tensorial structure could be developed is in place.     

A more specific discussion bring us back to section \ref{sec:GSusy}. Upon the use of generalized parallelizations to provide a frame for the generalized metric the path followed raises some issues. First, consider again the use we made of the embedding of $AdS_4$ in $\mathbb{R}^{2,3}$ to find $E^{a}$. Are there other ways of accomplishing the same goal and, on the other hand, can we extend this approach to different embeddings? For example, it would appear possible to treat in this fashion any quadric surface of dimension $4$ in $\mathbb{R}^{p,q}, p+q=5$ provided there is a volume-form, but it seems more difficult to treat non-orientable manifolds or embeddings of lower dimension. Remembering a manifold is endowed with a volume form if and only if it is orientable this would suggest an equivalence relation between $M$ being $d$-parallelizable and having an orientation (recall a parallelizable manifold is always orientable but the converse is not true \cite{Bishop}).  

Finally, we address ourselves to the maximal symmetry characteristic of our $4d$-supergravity solutions. By definition these spaces will have $\frac{d(d+1)}{2}=10$ Killing vector fields and we would like to know how this information is encoded when considering the generalized Killing vectors introduced in section \ref{sec:genK2}. More than that, it would be interesting to compare these sections with the basis constructed for $E^2$ (also of dimension $10$) using the $0$-parallelizations so as to study the existence of a connection between a $2$-global frame and the set of generalized Killing vector fields.
   
\appendix

\chapter{Notation and Conventions} \label{Ap:Notation}
Although the majority of the notation used is introduced throughout the text, for the sake of clarity we decided to also include a summary of it along with the conventions in use. We have compiled the following list:  
\begin{list}{-}{}
\item The signature of a metric space is denoted $(p,q)$ with $p$ positive and $q$ negative eigenvalues. We opted by the $(1,3)$ signature for Minkowski space;
\item A generalized vector is denoted $x=X+\xi$ with $X \in TM$ and $\xi \in \wedge^p T^{\ast}M$. The nomenclature for forms alone might vary: depending on the context they may be represented by lower case Greek letters ($\xi, \eta$) or upper case Latin letters ($B,C$); 
\item Latin letters from the end of the alphabet $(m,n,p,...)$ are used to denote generalized indices, as well as upper case ones $(A,B,...)$; Greek letters are reserved for space-time indices and Latin letters $(i,j,k,...)$ only for spacial ones and they will usually run from $1$ to $4$ and $2$ to $4$, respectively; Latin letters from the beginning of the alphabet $(a,b,c,...)$ are chosen to designate tangent space or generalized tangent space indices, although this is not so rigid;
\item Global frames and generalized global frames are distinguished by the use of lower and upper case, respectively e.g. $e,E$;
\item To avoid confusion, the determinant of a metric $g$ will always be denoted by $\tilde{g}$;
\item To make the difference between the Courant bracket and the usual Lie bracket always apparent they are denoted respectively using $\llbracket \cdot , \cdot \rrbracket $ and $ \left[ \cdot , \cdot \right]$ following Sheng \cite{Sheng}.    

\end{list}

\chapter{The isomorphism $\wedge^{p}T^{\ast}M\simeq(\det TM)\otimes\wedge^{d-p}T^{\ast}M$} \label{Ap:A}

On our effort to describe $4d$-supergravity using generalized geometry it will be more natural to treat $x=X + \xi$ in a representation where both vector field and form have components with ``indices upstairs". To do that we will use the isomorphism $\wedge^{p}T^{\ast}M\simeq(\det TM)\otimes\wedge^{d-p}T^{\ast}M$, with $d=\mbox{dim }M$. 

If $M$ is a manifold doted with a metric $g$ there is yet another interpretation for this isomorphism. Remember, by definition, the Hodge dual of a $p$-form $\xi = \frac{1}{p!} \xi_{\mu_1 \cdots \mu_p}dx^{\mu_1} \wedge \cdots \wedge dx^{\mu_p}$ is given by (\cite{Nakahara}, $p.290$):
{\begin{align}
\star \xi & = \frac{|\tilde{g}|^{\frac{1}{2}}}{p!(d-p)!} \xi_{\mu_1 \cdots \mu_p} \epsilon^{\mu_1 \cdots \mu_p}_{\mu_{p+1} \cdots \mu_{d}} dx^{\mu_{p+1}} \wedge \cdots \wedge dx^{\mu_d} \\
& = \frac{|\tilde{g}|^{\frac{1}{2}}}{(d-p)!} (\frac{1}{p!}\epsilon^{\nu_1 \cdots \nu_{d-p} \mu_1 \cdots \mu_p}\xi_{\mu_1 \cdots \mu_p})g_{\nu_1 \mu_{p+1}} \cdots g_{\nu_{d-p} \mu_m} dx^{\mu_{p+1}} \wedge \cdots \wedge dx^{\mu_d}. \nonumber
\end{align}}
Our (inverse) metric also allows us to ``raise indices" such that:
\begin{align}
(\star \xi)^{\ast} = \frac{|\tilde{g}|^{\frac{1}{2}}}{(d-p)!} (\frac{1}{p!}\epsilon^{\nu_1 \cdots \nu_{d-p} \mu_1 \cdots \mu_p}\xi_{\mu_1 \cdots \mu_p}) \frac{\partial}{\partial x^{\mu_{p+1}}} \wedge \cdots \wedge \frac{\partial}{\partial x^{\mu_{d}}}.  
\end{align}
So, apart from an overall factor we are actually dealing with the (metric) dual of a Hodge dual: an antisymmetric $(d-p,0)$-tensor. Hence, $\wedge^{p}T^{\ast}M\simeq(\det TM)\otimes\wedge^{d-p}TM$.  However, the isomorphism we started with is valid independently of the existence of a metric providing us, in any case, with a viable representation given by: 
\begin{align}
x^{\nu_1 \cdots \nu_{d-p}} & \equiv \frac{1}{p!} \epsilon^{\nu_1 \cdots \nu_{d-p} \mu_1 \cdots \mu_p} \xi_{\mu_1 \cdots \mu_p} \\
\Leftrightarrow \xi_{\mu_1 \cdots \mu_p} & = \frac{1}{(d-p)!} \epsilon_{\nu_1 \cdots \nu_{d-p} \mu_1 \cdots \mu_p} x^{\nu_1 \cdots \nu_{d-p}}.
\end{align}
We would now like to to use this representation to express usual operations involving $p$-forms. We begin with the interior product ($Y \in TM$),
{\begin{align}
i_Y \colon & \wedge^p T^{\ast}M \to (\det TM) \otimes \wedge^{d-p+1} T^{\ast}M \nonumber \\
& \frac{1}{p!} \xi_{\mu_1 \cdots \mu_p} \mapsto (i_Y x)^{\nu_1 \cdots \nu_{d-p+1}}= \frac{(-1)^{d-1}}{d-p} x^{[ \nu_1\cdots \nu_{d-p}} Y^{\nu_{d-p+1}]}. 
\end{align}}
As for the exterior derivative:
{\begin{align}
d \colon & \wedge^p T^{\ast}M \to (\det TM) \otimes \wedge^{d-p-1} T^{\ast}M \nonumber \\
& \frac{1}{p!} \xi_{\mu_1 \cdots \mu_p} \mapsto (dx)^{\nu_1 \cdots \nu_{d-p-1}}= (-1)^{d-p-1} \partial_{\mu} x^{[ \mu \nu_1\cdots \nu_{d-p-1}]}. 
\end{align}}
Since it will prove to be useful when treating the Courant bracket we shall also introduce the new representation for $d i_{Y}$:
{\begin{align}
d i_{Y} \colon & \wedge^p T^{\ast}M \to (\det TM) \otimes \wedge^{d-p} T^{\ast}M \nonumber \\
& \frac{1}{p!} \xi_{\mu_1 \cdots \mu_p} \mapsto (d i_{Y} x)^{\nu_1 \cdots \nu_{d-p}}= (d-p+1) \partial_{\mu} (Y^{[\mu}x^{\nu_1\cdots \nu_{d-p}]}). 
\end{align}}
Finally, we have the Lie derivative:
{\begin{align}
\mathcal{L}_{Y} \colon & \wedge^p T^{\ast}M \to (\det TM) \otimes \wedge^{d-p} T^{\ast}M \nonumber \\
& \frac{1}{p!} \xi_{\mu_1 \cdots \mu_p} \mapsto (\mathcal{L}_{Y} x)^{\nu_1 \cdots \nu_{d-p}}= Y^{\mu} \partial_{\mu}(x^{\nu_1\cdots \nu_{d-p}}) + (d-p+1)(\partial_{\mu}Y^{[\mu})x^{\nu_1\cdots \nu_{d-p}]}. 
\end{align}}

\chapter{Spinors in the generalized tangent bundle} \label{Ap:B}

 We define a \emph{Clifford algebra} $\mathpzc{Cl}(V)$, where $V$ is a vector space equipped with a quadratic form $F$, as an algebra with the following properties:
\begin{list}{$\bullet$}{}  
\item If $\lbrace e_i \rbrace_{i=1,\cdots,n}$ is a basis for $V$ then is a basis of generators for $\mathpzc{Cl}(V)$;
\item $(cd)e =c(de), \forall_{c,d,e \in \mathpzc{Cl}(V)}$ and $\exists_{1 \in \mathpzc{Cl}(V)} \forall_{c \in \mathpzc{Cl}(V)} \colon 1c = c1 = c$;
\item $v^2 = F(v), \forall_{v \in V} $.  
\end{list}
For $V \equiv E$, $F(\cdot)$ will be the quadratic form associated with the inner product $\langle \cdot , \cdot \rangle$, which is to say the last defining requirement becomes:
\begin{align}
x^2=\langle x,x \rangle, \forall_{x \in E}\mbox{ } \Leftrightarrow \mbox{ } xy + yx = 2 \langle x,y \rangle, \forall_{x,y \in E} 
\end{align}
Consider the exterior algebra for $T^{\ast}M$, $S = \wedge^{\bullet} T^{\ast}M \equiv \oplus_{p=0}^{m} \wedge^p T^{\ast}M$. It is isomorphic to $\mathpzc{Cl}(TM) \simeq \mathpzc{Cl}(T^{\ast}M)$, since $\langle X,Y\rangle =0, \forall_{X,Y \in TM}$ \cite{GualtieriLN} and it sits naturally inside $\mathpzc{Cl}(E)$. To see how, we will first introduce an action of $E$ over $S$,
\begin{align}
(\cdot) \colon & E \times S \to S \nonumber \\
& (X + \xi,\psi) \mapsto (X + \xi) \cdot \psi \equiv i_X \psi + \xi \wedge \psi,
\end{align}
that respects the defining property of our $\mathpzc{Cl}(E)$. In detail (\cite{GualtieriPhD}, $p.8$):
\begin{align}
(X + \xi) \cdot (X + \xi,\psi)) & = i_X (i_X \psi + \xi \wedge \psi)+ \xi \wedge(i_X \psi + \xi \wedge \psi) \nonumber \\ 
& = (i_X \xi) \wedge \psi - \xi \wedge (i_X \psi) + \xi \wedge (i_X \psi) + (\xi \wedge \xi) \wedge \psi \mbox{, using } i_{X}^2=0 \nonumber \\
& = \xi(X)\psi \mbox{, since } \xi \wedge \xi =0 \mbox{ and } i_X \xi =\xi (X) \in C^{\infty}(M) \mbox{ for }\xi \in T^{\ast}M \nonumber \\
& = \langle X + \xi, X + \xi \rangle \psi.
\end{align}
This action induces a representation $(\cdot)$ between $\mathpzc{Cl}(E)$ and $S$, referred to as the \emph{spin representation}:
\begin{align}
(\cdot) \colon & \mathpzc{Cl}(E) \times S \to S \nonumber \\
& (a1 + a_i{\tilde{e}}^{i}+ \cdots + a_{i_1 \cdots i_{2d}}{\tilde{e}}^{i_1} \cdots {\tilde{e}}^{i_{2d}}, \psi ) \mapsto (a1 + a_i{\tilde{e}}^{i}+ \cdots + a_{i_1 \cdots i_{2d}}{\tilde{e}}^{i_1} \cdot ... \cdot{\tilde{e}}^{i_{2d}}) \cdot \psi, 
\end{align}
where $\lbrace {\tilde{e}}^{i} \rbrace_{i=1,\cdots,2d}$ is a basis for $TM \oplus T^{\ast}M$. 

We see how the exterior algebra provided us with a representation for spinors: unfortunately, this approach does not carry on for the generalized bundle $TM \oplus \wedge^2 T^{\ast}M$ since the action described before does not respect the condition $(X + \xi)^2 \cdot \psi = \langle X + \xi, X + \xi \rangle $ for $\xi \in \wedge^2 T^{\ast}M$.

The spin group $Spin(E)$ is defined as
\begin{align}
Spin(E) \equiv \left\lbrace x_1 \cdots x_r, r \mbox{ even } \colon x_i \in E, \langle x_i,x_i \rangle= \pm 1 \right\rbrace,
\end{align}
and provides a double cover for $SO(E)$ via an homomorphism that can be represented on $E$ by the adjoint action:
\begin{align}
\mathrm{Ad} \colon & Spin(E) \times E \to SO(E) \times E \nonumber \\
& (s,x) \mapsto \mathrm{Ad}_s(x) \equiv s x s^{-1}.
\end{align}
We have mentioned in section \ref{sec:Elinear} the identity $\mathfrak{so}(E)=\wedge^2 E$, meaning the elements of this Lie algebra are naturally included in $\mathpzc{Cl}(TM) \subset \mathpzc{Cl}(E)$. We will be able to see this explicitly by means of the representation induced by $\mathrm{Ad}$
\begin{align}
\mathrm{ad} & \colon \mathfrak{so}(E) \times E \to E \nonumber \\
& (Q,x) \mapsto \mathrm{ad}_Q (x) \equiv [Q,x],
\end{align}
and using the isomorphism $\wedge^2 E \simeq \wedge^2 T^{\ast}M \oplus (TM \otimes T^{\ast}M) \oplus \wedge^2 TM$, i.e, the decomposition $Q = B + A + \beta$. Let $\lbrace e_{i}\rbrace_{i=1,\cdots,d}$ and $\lbrace e^{i}\rbrace_{i=1,\cdots,d}$ denote basis for $TM$ and $T^{\ast}M$, respectively: we can write $B= \frac{1}{2} B_{ij}e^{i} \wedge e^{j}$, $A=A_j^{i} e_{i} \otimes e^j$ and $\beta =\frac{1}{2} \beta^{ij} e_i \wedge e_j$ and use this to investigate how can they be viewed as elements of $\mathpzc{Cl}(E)$. With that purpose in mind we introduce a basis for the generators of the Clifford algebra given by the $O(d,d)$ gamma matrices (\cite{GranaFlux}, $p.19$):
\begin{align}
\lbrace \tilde{e}^{i} & =\Gamma^{i} \equiv e^{i} \wedge \rbrace_{i=1,\cdots ,d} \\
\lbrace \tilde{e}_{i} & =\Gamma_{i} \equiv i_{e_i} \rbrace_{i=1,\cdots ,d}
\end{align}
Under this representation we have:
\begin{align}
\mathrm{ad}_{e^{i} \wedge e^{j}}(\cdot) \colon & e_{i} \mapsto [e^{i} \wedge e^{j},e_{i}]
= e^{i}(e_{i}) \wedge e^{j} + e^{i} \wedge e^{j} (e_{i}) - e_{i}(e^{i})\wedge e^{j} 
= e^{i} \delta_{i}^{j} = e^{j} \nonumber \\
{\Gamma}^{j} {\Gamma}^{i} \colon & e_{i} \mapsto ({\Gamma}^{j} {\Gamma}^{i})(e_{i})= e^{j} \nonumber \\
\Rightarrow  B  = \frac{1}{2} & B_{ij}{\Gamma}^{j} {\Gamma}^{i} \\
\mathrm{ad}_{e_{i} \otimes e^{j}}(\cdot) \colon & e_{j} \mapsto [e_{i} \otimes e^{j},e_{j}]
=e_{j}(e_{i} \otimes e^{j})+ e_{i} \otimes (e^{j}e_{j})- e_{j}(e_{i} \otimes e^{j})=e_i \nonumber \\
{\Gamma}_{i} {\Gamma}^{j} \colon & e_{j} \mapsto ({\Gamma}_{i} {\Gamma}^{j})(e_{j})= e_{i} \nonumber \\
\mathrm{ad}_{e_{i} \otimes e^{j}}(\cdot) \colon & e^{i} \mapsto [e_{i} \otimes e^{j},e^{i}]
=-e^{j} \otimes (e_{i}e^{i}) = -e^{j} \nonumber \\
-{\Gamma}^{j} {\Gamma}_{i} \colon & e^{i} \mapsto -({\Gamma}^{j} {\Gamma}_{i})(e^{i})=- e^{j} \nonumber \\
\Rightarrow  A  = \frac{1}{2} & A_{j}^{i}({\Gamma}_{i} {\Gamma}^{j}-{\Gamma}^{j} {\Gamma}_{i}) \\
\mathrm{ad}_{e_{i} \wedge e_{j}}(\cdot) \colon & e^{i} \mapsto [e_{i} \wedge e_{j},e^{i}]
= e_{i}(e^{i}) \wedge e_{j} + e_{i} \wedge e_{j} (e^{i}) - e^{i}(e_{i})\wedge e_{j} 
= e_{i} \delta^{i}_{j} = e_{j} \nonumber \\
{\Gamma}_{j} {\Gamma}_{i} \colon & e^{i} \mapsto ({\Gamma}_{j} {\Gamma}_{i})(e^{i})= e_{j} \nonumber \\
\Rightarrow  \beta  = \frac{1}{2} & \beta^{ij}{\Gamma}_{j} {\Gamma}_{i} .
\end{align}
This allow us to see how $\mathfrak{so}(E)$ acts under the spin representation, and moreover, by exponentiation, we can also learn the behaviour of the $SO(E)$ elements (\cite{GualtieriPhD}, $p.9$):  
\begin{align}
B \cdot \psi & = \frac{1}{2}B_{ij} {\Gamma}^{j}{\Gamma}^{i}\psi = \frac{1}{2}B_{ij} {e}^{j} \wedge {e}^{i} \wedge \psi = -B \wedge \psi \\
\Rightarrow e^B \cdot \psi & = (1-B+\frac{1}{2}B \wedge B + \cdots) \wedge \psi
\end{align}
\begin{align} 
A \cdot \psi & = \frac{1}{2} A^{i}_{j}({\Gamma}_{i}{\Gamma}^{j}-{\Gamma}^{j}{\Gamma}_{i})\cdot \psi 
 = \frac{1}{2} A^{i}_{j}(e_{i} \cdot (e^{j} \wedge \psi)- e^{j}(i_{e_i} \psi)) \\
& = \frac{1}{2} A^{i}_{j}((i_{e_i}e^{j}) \wedge \psi - 2 e^{j} \wedge (i_{e_i} \psi)) 
 =\frac{1}{2}(A^{i}_{j}\delta_{i}^{j}-2A^{tr})\psi = \frac{1}{2} \mbox{Tr }A\psi -A^{tr}\psi, \nonumber \\
 & \mbox{with} \\
 A \colon & \wedge^p TM \to \wedge^p TM \nonumber \\
& \frac{1}{p!} V^{i_1 \cdots i_p} \mapsto (A V)^{i_1 \cdots i_p}
= \frac{1}{p!} A^{i_1}_{j}V^{j i_2 \cdots i_p} \nonumber \\
 \Rightarrow e^{A} \cdot \psi & = (\det e^A)^{\frac{1}{2}}(e^{-A})^{tr} \\
\beta \cdot \psi & = \frac{1}{2}\beta^{ij} {\Gamma}_{j}{\Gamma}_{i} \psi =	 \frac{1}{2}\beta^{ij} i_{e_{j}} (i_{e_{i}} \psi) =i_{\beta} \psi \\
& \mbox{with the definition } \nonumber \\
i_{\beta} \colon & \wedge^p TM \to \wedge^{p-2} TM \nonumber \\
& \frac{1}{p!}\xi_{k_1 \cdots k_p} \mapsto i_{\beta}\xi = \frac{1}{p!} ( \frac{1}{2} \beta^{ij} ) \xi_{ijk_3 \cdots k_{p}} \nonumber \\
\Rightarrow e^{\beta} \cdot  \psi & = e^{i_{\beta}} \psi.
\end{align}

As it happened for the case of  $GL^{+}(TM)$ where the embedding could be extended to the whole of $GL(TM)$ the same is also true for the spin representation: details of the explicit construction can be found in \cite{GualtieriPhD} ($p.10$).

The representation provided by $S$ for the spin group is not irreducible. We find one by considering the (oriented) \emph{volume element} of the Clifford algebra. Following  the conventions in \cite{Lawson} ($p.21$) it is defined as $\omega =\hat{e}_1^{-}\cdots \hat{e}_d^{-}\hat{e}_1^{+}\cdots\hat{e}_d^{+}$, where, we note, $\lbrace \hat{e}_i^{\pm} = \Gamma_i \pm \Gamma^i \rbrace_{i=1,\cdots,d}$ satisfies $\langle \hat{e}_i^{\pm}, \hat{e}_j^{\pm} \rangle \psi = \pm \delta_{ij} \psi$ and $\langle \hat{e}_i^{\pm}, \hat{e}_j^{\mp} \rangle \psi = 0$, $\forall_{\psi \in S}$ (\cite{GualtieriLN}, $p.11$). Using the properties of this basis we prove:
\begin{align}
\omega^2 & = \hat{e}_1^{-}\cdots \hat{e}_d^{-}\hat{e}_1^{+}\cdots\hat{e}_d^{+} 
\hat{e}_1^{-}\cdots \hat{e}_d^{-}\hat{e}_1^{+}\cdots\hat{e}_d^{+} \nonumber \\
& =(-1)^{2d-1}\hat{e}_1^{-}\cdots (-1)^{d}\hat{e}_d^{-}\hat{e}_d^{-}
(-1)^{d-1}\hat{e}_1^{+}\hat{e}_1^{+}\cdots \hat{e}_d^{+}\hat{e}_d^{+} \nonumber \\
& = (-1)^{d(2d-1)+d}.1 \nonumber \\
& = 1.
\end{align}
We can then define a partition of the exterior algebra on the $+1$ and $-1$ eigenspaces of $\omega$: $S\equiv S^{+} \oplus S^{-}$. It remains to identify what the spaces $S^{\pm}$ correspond to. We then must analyse how $\omega$ acts on a general element of $S$, using the action defined earlier. Let us have a particular case in mind: $\psi=e^{1} \wedge \cdots \wedge e^{p} \in S$. The action of $\omega$ brings $\psi$ to:
\begin{align}
 \omega \psi &= \hat{e}_1^{-} \cdots \hat{e}_d^{-}\hat{e}_1^{+}\cdots \hat{e}_{d-1}^{+} \cdot (i_{e_{d}}+e^{d} \wedge)(e^{1}\wedge \cdots \wedge e^{p}) \nonumber \\
 & = (-1)^{p \left(d - \frac{p+1}{2} \right)}\hat{e}_1^{-} \cdots \hat{e}_d^{-}\hat{e}_1^{+}\cdots \hat{e}_{p}^{+} \cdot (e^{p}\wedge \cdots \wedge e^{1} \wedge e^{p+1} \wedge \cdots \wedge e^{m}) \nonumber \\
 & = (-1)^{\frac{d(d-1)}{2}}\hat{e}_1^{-} \cdots \hat{e}_d^{-} \cdot(e^{p+1} \wedge \cdots \wedge e^{d}) \nonumber \\
 & = (-1)^{\frac{d(d-1)}{2}+ p} \psi.
\end{align}
The procedure is readily generalizable for any element of the exterior algebra and, by linearity, holds the same result. We find $S^{\pm}$ can be identified with $\wedge^{even/odd} T^{\ast}M \equiv \oplus_{p=0} \wedge^{2p/2p+1}T^{\ast}M$ depending on the dimension of the manifold. Specifically:
\begin{align*}
d  & = 0,1 \pmod{4} \Rightarrow S^{+/-}=\wedge^{even/odd} T^{\ast}M \\
d  & = 2,3 \pmod{4} \Rightarrow S^{+/-}=\wedge^{odd/even} T^{\ast}M.
\end{align*} 
If we want the equality $S^{+/-}=\wedge^{even/odd} T^{\ast}M$ to hold for all $d$ we need only to make a redefinition of our volume element $\omega \to(-1)^{\frac{d(d-1)}{2}} \omega$ as in \cite{GualtieriLN}, $p.16$. 

\chapter{Spin structure for $E^2$} \label{Ap:C}

The construction of a Clifford algebra for $E^2$ does not follow directly from the one done for $E$\footnote{See Appendix C}, where we would see $O(TM)$ spinors arise naturally as a subgroup of $Spin(d,d)$. Nevertheless, from the treatment developed in section \ref{sec:TMlinear} we know $GL(TM)$ to be included in $GL(E^2)$ so naturally the spin representations of interest arise in our formalism as well. We certainly would still expect to find $\mathpzc{Cl}(TM) \subset \mathpzc{Cl}(E^2)$ (albeit the latter has not been defined) meaning that for $M=\mathbb{R}^{1,3}$ we would have $Spin(1,3)$ sitting in $\mathpzc{Cl}(T\mathbb{R}^{1,3}) \equiv \mathpzc{Cl}(1,3)$ included in the more general Clifford algebra. 

The inclusion of the whole algebra $\mathpzc{Cl}(1,3)$ will bring along an additional spin group, $Spin(2,3)$. The matrix its Lie algebra elements preserve will provide us with a motivation for the choice of the signature for the generalized metric, the same the maximally symmetric solutions for \eqref{eq:metric} already gave us: $(2,3)$.

Consider then the Clifford algebra $\mathpzc{Cl}(1,3)\simeq GL(4,\mathbb{R})$, generated by the (real) antisymmetric combinations of gamma matrices, i.e., with generators $\lbrace \mathds{1},\gamma^{\mu},\gamma^{\mu_1 \mu_2},\gamma^{\mu_1 \mu_2 \mu_3},\gamma^{\mu_1 \cdots \mu_4} \rbrace$. We would like to find a basis for the Lie algebra of $Spin(2,3) \simeq Sp(2,\mathbb{R}) \subset \mathpzc{Cl}(1,3)$, as we have for $Spin(1,3)$, where the ${\gamma^{\mu_1 \mu_2}}$ play that role, in particular, making the relation $Spin(1,3) \subset Spin(2,3)$ apparent.

We start by using the exceptional isomorphism referred. The defining property of the symplectic group $Sp(2,\mathbb{R})$ is as follows (\cite{Hall}, $p.$):
\begin{align}
Sp(2,\mathbb{R})= \lbrace A \in {\mathbb{M}}_{4�4}(\mathbb{R})\colon A \Omega A^{tr}= \Omega \rbrace ,\Omega^{tr}=-\Omega.
\end{align}
From this definition we obtain the Lie algebra of the group:
\begin{align}
\mathfrak{sp}(2,\mathbb{R})= \lbrace X \in {\mathbb{M}}_{4�4}(\mathbb{R})\colon (\Omega X)^{tr}=\Omega X \rbrace.
\end{align}
We remember that a basis for a set of symmetric matrices $\lbrace T^{\alpha \beta} \rbrace$ can be given by $(T^{\alpha \beta})_{\mu \nu} = \delta_{\mu}^{\alpha} \delta_{\nu}^{\beta} + \delta_{\nu}^{\alpha} \delta_{\mu}^{\beta}$. Our Lie algebra basis ${X^{\alpha \beta}}_{1 \leq \alpha \leq \beta \leq 4}$ should then satisfy:
\begin{align}
(X^{\alpha \beta})_{\nu}^{\sigma}=(\Omega^{-1})^{\sigma \alpha}\delta_{\nu}^{\beta}+(\Omega^{-1})^{\sigma \beta}\delta_{\nu}^{\alpha}.
\end{align}
We would now like to express our basis in terms of the antisymmetric combinations of gamma matrices. For that, we will choose a convenient $\Omega$:
\begin{align}
\Omega = \left( \begin{smallmatrix}
  0  & 1 &  0 & 0 \\
 -1  & 0 &  0 & 0 \\
  0  & 0 &  0 & 1 \\
  0  & 0 & -1 & 0 \\
\end{smallmatrix} \right).
\end{align}
Our basis, written in terms of gamma matrices, is then:
\begin{align*}
X^{11} & = -\frac{1}{2} \gamma^{13}+\frac{i}{2} \gamma^{23}-\frac{1}{2} \gamma^{013}+\frac{i}{2} \gamma^{023},\\
X^{22} & = -\frac{1}{2} \gamma^{13}-\frac{i}{2} \gamma^{23}-\frac{1}{2} \gamma^{013}-\frac{i}{2} \gamma^{023},\\
X^{33} & = -\frac{1}{2} \gamma^{13}+\frac{i}{2} \gamma^{23}+\frac{1}{2} \gamma^{013}-\frac{i}{2} \gamma^{023},\\
X^{44} & = -\frac{1}{2} \gamma^{13}-\frac{i}{2} \gamma^{23}+\frac{1}{2} \gamma^{013}+\frac{i}{2} \gamma^{023},\\
X^{12} & = -\frac{i}{2} \gamma^{12}-\frac{i}{2} \gamma^{012},\\
X^{34} & = -\frac{i}{2} \gamma^{12}+\frac{i}{2} \gamma^{012},\\
X^{13} & =  \frac{1}{2} \gamma^{01}-\frac{i}{2} \gamma^{02}, \\
X^{24} & = -\frac{1}{2} \gamma^{01}-\frac{i}{2} \gamma^{02}, \\
X^{14} & = -\frac{1}{2} \gamma^{03}+\frac{i}{2} \gamma^{123},\\ 
X^{23} & = -\frac{1}{2} \gamma^{03}-\frac{i}{2} \gamma^{123},\\
\end{align*}
which means that a basis for the Lie algebra in question might be $\lbrace \gamma^{\mu_1 \mu_2},\gamma^{\mu_1 \mu_2 \mu_3} \rbrace$\footnote{We note however this result is not independent of the matrix $\Omega$ chosen: for $\Omega '=S \Omega S^{-1}$ the basis for the Lie algebra would appear transformed by the same similarity transformation. The two groups and respective Lie algebras are of course isomorphic but one must be careful when choosing the representative antisymmetric matrix simply to ensure the result comes in a standard format.}.   

We expect it possible to express this new basis as a set o gamma matrices itself, in the sense its elements $\Gamma^m,m=0,\cdots,4$ satisfy $\lbrace \Gamma^{m},\Gamma^{n} \rbrace =2\eta^{mn}\mathds{1}$, $\eta=\mbox{diag}(1,-1,-1,-1,1)$. It turns out we encounter such a set $\lbrace \Gamma^{mn}=\Gamma^{[m}\Gamma^{n]} \rbrace_{m,n=0,\cdots,4}$ to be exactly the one presented before, with the identification $\Gamma^{\mu}=\gamma^{\mu},\mu=0,\cdots,3$ and $\Gamma^4=-i\gamma^{0} \gamma^{1} \gamma^{2} \gamma^{3}$. 
  
\chapter{Courant bracket for $E^2$: Details on the Representations} \label{Ap:D}

Let us consider the dual space $(E^2)^{\ast}=T^{\ast}M \oplus \wedge^{2}TM$. We would like to introduce an operator $\partial_{mn} \in (E^2)^{\ast}$, that generalizes $\partial_{\mu} \in T^{\ast}M$. We can think immediately of three possible generalizations:
\[
(\partial)_{mn} \equiv \left( \begin{array}{cc} 0 & 0 \\ \partial_{\mu} & 0 \end{array} \right) \mbox{ , }
(\partial^{S})_{mn} \equiv (\partial)_{mn}+(\partial)_{mn}^{tr} \mbox{ , }
(\partial^{A})_{mn} \equiv (\partial)_{mn}-(\partial)_{mn}^{tr}. 
\]
We begin by seeing what is brought to us by the combinations $x^{mp}\partial_{pq}y^{qn}$ and $x^{mp}\partial_{pq}^{tr}y^{qn}$:
\begin{align}
{\left( x \partial y \right)}^{mn} 
& = \left( \begin{array}{cc} -x^{5 \mu_1}\partial_{\nu}y^{\nu \mu_2} & x^{5 \mu_1}\partial_{\nu}y^{5 \nu}\\
0 & 0 \end{array} \right) \\
{\left( x \partial^{tr} y \right)}^{mn} 
& = \left( \begin{array}{cc} x^{\mu_1 \nu}\partial_{\nu}y^{5 \mu_2} & 0 \nonumber \\
x^{5 \nu}\partial_{\nu}y^{5 \mu_2} & 0 \end{array} \right). 
\end{align}
Now since the weighted cyclic permutations of $x^{mp}\partial_{pq}^{tr}y^{qn}$ give us:
\begin{align}
x^{[m|p}\partial_{pq}^{tr}y^{q|n]} + \mbox{w.c.p.}(p,q,[n,m])= \left( \begin{array}{cc}
-x^{5 \nu}\partial_{\nu}y^{\mu_1 \mu_2} + \mbox{w.c.p.}(5,\nu,[\mu_1,\mu_2]) & 0 \\
2x^{5 \nu}\partial_{\nu}y^{5 \mu_2} & 0
\end{array} \right), 
\end{align}
it makes sense to consider the combination $y^{[m|p}\partial_{pq}^{tr}x^{q|n]}-x^{[m|p}\partial_{pq}^{tr}y^{q|n]}+ \mbox{w.c.p.}(p,q,[n,m])$, and from the above expression it follows directly:
\begin{align}
y^{[m|p}\partial_{pq}^{tr}x^{q|n]}-x^{[m|p}\partial_{pq}^{tr}y^{q|n]} + \mbox{w.c.p.}(p,q,[n,m]) =
\left( \begin{array}{cc}
\begin{array}{cr}
& x^{5 \nu}\partial_{\nu}y^{\mu_1 \mu_2} - y^{5 \nu}\partial_{\nu}x^{\mu_1 \mu_2} \\
& + \mbox{w.c.p.}(5,\nu,[\mu_1,\mu_2]) \end{array} & 0 \\
-2 \left( x^{5 \nu}\partial_{\nu}y^{5 \mu_2} - y^{5 \nu}\partial_{\nu}x^{5 \mu_2} \right) & 0 
\end{array} \right)
\end{align}
This brings us to the study of $\frac{1}{2}[(y^{[m|p}\partial_{pq}^{tr}x^{q|n]}-x^{[m|p}\partial_{pq}^{tr}y^{q|n]})-(y^{[m|p}\partial_{pq}^{tr}x^{q|n]}-x^{[m|p}\partial_{pq}^{tr}y^{q|n]})^{tr}]+$ w.c.p.$(p,q,[n,m])$, i.e., the antisymmetric combination of the preceding terms. One actually finds out that if we set $\alpha=-1$ the relation below holds:
\begin{align}
{ \llbracket x,y \rrbracket }^{mn} & = -\frac{1}{4}[(y^{[m|p}\partial_{pq}^{tr}x^{q|n]}-x^{[m|p}\partial_{pq}^{tr}y^{q|n]})-(y^{[m|p}\partial_{pq}^{tr}x^{q|n]}-x^{[m|p}\partial_{pq}^{tr}y^{q|n]})^{tr}]+\mbox{w.c.p.}(p,q,[n,m]) \nonumber \\
& = \frac{1}{4}[x^{[m|p}\partial_{pq}^{A}y^{q|n]}-y^{[m|p}\partial_{pq}^{A}x^{q|n]}]+\mbox{w.c.p.}(p,q,[n,m]).
\end{align}
This suggests we should set the generalization of $\partial_{\mu}$ as $\partial_{mn}^{A}$.

\chapter{$AdS_4$ in hyperbolic coordinates} \label{Ap:E} 

We consider a diffeomorphism for $\mathbb{R}^{2,3}$ that takes Cartesian coordinates $\lbrace x^{m}\rbrace_{m=1,\cdots,5}$ to hyperbolic ones $ \lbrace R, \tau, \rho, \theta, \phi \rbrace$ such that $AdS_4$ can be seen as the surface defined by $R=\frac{3}{\lambda^2}$. The explicit coordinate transformation is given by:
\begin{align*}
x^1 &= R \cosh \rho \cos \tau ,\\
x^2 &= R \sinh \rho \cos \phi \sin \theta ,\\
x^3 &= R \sinh \rho \sin \phi \sin \theta ,\\
x^4 &= R \sinh \rho \cos \theta ,\\
x^5 &= R \cosh \rho \sin \tau.
\end{align*}
The metric induced by this diffeomorphism was presented in section \ref{sec:GSusy}. Here we present the Jacobian associated with this transformation, so that the similarities between the generalized parallelization in terms of one-form and volume-form become apparent, again as introduced in  section \ref{sec:GSusy}:
\begin{align}
J_f &= \left( \frac{\partial x^m}{\partial z^n}\right)
&= \left( \begin{smallmatrix}
\cosh \rho \cos \tau & R \sinh \rho \cos \tau & 0 & 0 & -R\cosh \rho \sin \tau \\
\sinh \rho \cos \phi \sin \theta & R \cosh \rho \cos \phi \sin \theta & R \sinh \rho \cos \phi \cos \theta 
& -R \sinh \rho \sin \phi \sin \theta & 0 \\
\sinh \rho \sin \phi \sin \theta & R \cosh \rho \sin \phi \sin \theta & -R \sinh \rho \sin \phi \cos \theta 
& R \sinh \rho \cos \phi \sin \theta & 0 \\
\sinh \rho \cos \theta & R \cosh \rho \cos \theta & -R \sinh \rho \sin \theta & 0 & 0 \\
\cosh \rho \sin \tau & R \sinh \rho \sin \tau & 0 & 0 & R \cosh \rho \cos \tau
\end{smallmatrix} \right).
\end{align}

\end{document}